# Current-driven collective control of helical spin texture in van der Waals antiferromagnet

Kai-Xuan Zhang,[1,2†*$] Suik Cheon,[3†] Hyuncheol Kim,[1,2†] Pyeongjae Park,[1,2¶] Yeochan An,[1,2] Suhan Son,[1,2‡] Jingyuan Cui,[1,2§] Jihoon Keum,[1,2] Joonyoung Choi,[4] Younjung Jo,[4] Hwiin Ju,[5] Jong-Seok Lee,[5] Youjin Lee[1,2#], Maxim Avdeev,[6,7] Armin Kleibert,[8] Hyun-Woo Lee,[3,9*] Je-Geun Park[1,2*]

[1]*Department of Physics and Astronomy & Institute of Applied Physics, Seoul National University, Seoul 08826, South Korea*
[2]*Center for Quantum Materials, Seoul National University, Seoul 08826, South Korea*
[3]*Department of Physics, Pohang University of Science and Technology, Pohang 37673, South Korea*
[4]*Department of Physics, Kyungpook National University, Daegu 41566, South Korea*
[5]*Department of Physics and Photon Science, Gwangju Institute of Science and Technology (GIST), Gwangju 61005, South Korea*
[6]*Australian Nuclear Science and Technology Organization, Locked Bag 2001, Kirrawee DC, NSW 2232, Australia*
[7]*School of Chemistry, The University of Sydney, Sydney, NSW 2006, Australia*
[8]*Swiss Light Source, Paul Scherrer Institut, Villigen PSI CH-5232, Switzerland*
[9]*Asia Pacific Center for Theoretical Physics, Pohang 37673, South Korea*
[¶]*Present address: Materials Science and Technology Division, Oak Ridge National Laboratory, Oak Ridge, Tennessee 37831, USA*
[‡]*Present address: Department of Physics, University of Michigan, Ann Arbor, MI 48109, USA*
[§]*Present address: Department of Physics & Astronomy, University of California, California, 92521, USA*
[$]*Present address: Department of Physics, Washington University in St. Louis, St. Louis, Missouri 63130, USA*
[#]*Present address: Center for Integrated Nanotechnologies, Los Alamos National Laboratory, Los Alamos, New Mexico 87545, USA*

[†] These authors contributed equally to this work.
[*] Corresponding authors: Kai-Xuan Zhang (email: kxzhang.research@gmail.com), Hyun-Woo Lee (email: hwl@postech.ac.kr), and Je-Geun Park (email: jgpark10@snu.ac.kr).

**Abstract**
Electrical control of quantum magnetic states is essential in spintronic science. Initial studies on the ferromagnetic state control were extended to collinear antiferromagnets and, more recently, noncollinear antiferromagnets. However, electrical control mechanisms of such exotic magnetic states remain poorly understood. Here, we report the first experimental and theoretical example of the current control of helical antiferromagnets, arising from the competition between collinear antiferromagnetic exchange and interlayer Dzyaloshinskii-Moriya interaction in new van-der-Waals (vdW) material $Ni_{1/3}NbS_2$. Due to the intrinsic broken inversion symmetry, an in-plane current generates spin-orbit torque that, in turn, interacts directly with the helical antiferromagnetic order. Our theoretical analyses indicate that a weak ferromagnetic order coexists due to the Dzyaloshinskii-Moriya interaction, mediating the spin-orbit torque to collectively rotate the helical antiferromagnetic order. Our $Ni_{1/3}NbS_2$ nanodevice experiments produce current-dependent resistance change consistent with the theoretical prediction. This work widens our understanding of the electrical control of helical antiferromagnets and promotes vdW quantum magnets as interesting material platforms for electrical control.

**Main Text**
**Introduction**
In spintronics, antiferromagnetic materials host various advantages like stray field elimination, ultrafast dynamics in the terahertz range [1], and stability to magnetic perturbations. These merits may be utilized to realize more compact, faster, and robust memory [2-4]. In this respect, collinear antiferromagnetic metals have received considerable attention [2-7]. Studies on the electrical control of collinear antiferromagnets are being extended to noncollinear antiferromagnets [8-10]. A recent experiment revealed that an electrical current applied to the Kagome semimetal $Mn_3Sn$ can switch its chiral antiferromagnetic order [11]. It was reported that the magnetic octupole coexisting with the chiral antiferromagnetic order plays an essential role in electrical control. However, the coupled dynamics of the chiral antiferromagnetic order and magnetic octupole is anomalous [12] and demands an improved understanding [13]. Another experiment [14] highlighted that an electrical current applied to Fe-intercalated van-der-Waals (vdW) material $Fe_{1/3}NbS_2$ demonstrated the electrical control at surprisingly low current densities of $10^4$ A/cm$^3$. Further investigations indicated that the spin-glass order [14,15] coexists in the material and regulates the electrical control. However, its mechanism again requires clarifications.

vdW materials provide excellent platforms to explore the intriguing quantum phenomena in these two-dimensional (2D) systems. Moreover, various vdW materials can be stacked together with high compatibility to build heterostructures and tailored with high tunability, making them promising components for device science [16-18]. These virtues recently attracted extensive attention and research interest in vdW materials. Specifically, the advent of magnetic vdW materials [19-23] has prompted a flurry of activities investigating 2D magnetism [24,25] and exploiting their new opportunities for spintronic physics [26-32]. On the other hand, the helical spin order is physically fascinating with its inborn helicity and many-body features. Unfortunately, it remains unknown whether a pure current can control such a unique type of spin texture, i.e., the helical Néel order in vdW antiferromagnets.

In this work, we investigate a vdW antiferromagnet, Ni-intercalated transition metal dichalcogenide $Ni_{1/3}NbS_2$ with a unique helical antiferromagnetic order [Fig. 1(f)]. We predict the electrical control of $Ni_{1/3}NbS_2$'s helical spin texture through macrospin simulations due to the current-driven antiferromagnetic spin-orbit torque. Further theoretical analysis suggests that the Dzyaloshinskii-Moriya interaction induces a weak ferromagnetic order [Fig. 2(a)] coexisting with a helical antiferromagnetic order. Finally, the former order is a handy knob to manipulate the entire helical antiferromagnetic spin texture electrically. Then, we experimentally demonstrate such current control in nanometer-thin $Ni_{1/3}NbS_2$ nanodevices. For example, the symmetry of the current-controlled resistance change in the experiment reinforces our scenario that the current drives the collective rotation of the helical order through the current-induced spin-orbit torque. Moreover, the resistance change exhibits a transition-like character near the Néel temperature in its temperature-current mapping, pointing to its magnetism and corresponding spin-orbit torque origin. Our work constitutes the first demonstration of the electrical control of the helical order.

**Results**

**Basic physical properties of $Ni_{1/3}NbS_2$**

As shown in Fig. 1(a), Ni atoms are intercalated between the vdW $NbS_2$ layers with a chiral screw *c*-axis, reducing the original $P6_3/mmc$ space group to the chiral space group $P6_322$ of $Ni_{1/3}NbS_2$. Fig. 1(b) exhibits the single-crystal x-ray diffraction (XRD) with well-resolved sharp peaks consistent with its hexagonal structure of the $P6_322$ space group. In addition, the powder XRD in Fig. 1(c) shows a strong (101) peak with an intensity much larger than that of the (100) and (102) peaks. Such a prominent (101) peak is the signature for the noncentrosymmetric space group $P6_322$ and the resultant inversion symmetry breaking [33]. The temperature-dependent susceptibility in Fig. 1(d) reflects the paramagnetic-to-antiferromagnetic transition near the Néel temperature $T_N$ of ~85 K. Subsequently, we performed the neutron powder diffraction experiment above (100 K) and below (5 K) the $T_N$ [Fig. 1(e)]. In contrast to the linear background at 100 K, the diffraction pattern at 5 K shows two satellite peaks equally split sideways from the (001) position in the inset, indicating the magnetic propagation vector of $Q_m = (0, 0, \delta)$ (r.l.u.) with $\delta \sim 0.03$ [34]. Therefore, $Ni_{1/3}NbS_2$ is an antiferromagnet with a helical spin structure along the *c*-axis, with a period of 66 layers. Simply put, $Ni_{1/3}NbS_2$ represents a 2D helical antiferromagnet with the in-plane spin texture of an A-type model [Fig. 1(f)]. We want to refer to this particular helical antiferromagnet as a "layered helical antiferromagnet": two helical spin chains intertwined mutually with neighboring layer spin being slightly noncollinear.

**Theoretical simulations on current control of helical orders**

Of further interest, nickel atoms' intercalation introduces in-plane antiferromagnetism and breaks global inversion symmetry, making $Ni_{1/3}NbS_2$ an ideal testbed to scrutinize spin-orbit torque effects on a helical antiferromagnet. To this end, we performed the macrospin simulations to predict the possible current control of its spin states by spin-orbit torque.

The spin-spin interaction in the helical antiferromagnet $Ni_{1/3}NbS_2$ consists of an intralayer ferromagnetic exchange interaction and an interlayer antiferromagnetic exchange interaction. For ease of discussion, we simplify the helical antiferromagnet to be a one-dimensional local spin chain model along the *c*-axis (*z*-axis) with the spin Hamiltonian:

$$H_{spin} = \sum_i^N [\hbar\omega_{ex} \mathbf{S}_i \cdot \mathbf{S}_{i+1} + \hbar\omega_{DMI} \hat{\mathbf{z}} \cdot (\mathbf{S}_i \times \mathbf{S}_{i+1}) + \hbar\omega_z(\mathbf{S}_i \cdot \hat{\mathbf{z}})^2 - \hbar\omega_{in}(\mathbf{S}_i \cdot \boldsymbol{\varphi}_0)^2], \qquad (1)$$

where $N$ is the number of magnetic layers, $\mathbf{S}_i$ the spin at layer $i$, $\hbar\omega_{ex}$ the interlayer antiferromagnetic spin exchange interaction strength, $\hbar\omega_{DMI}$ the interlayer Dzyaloshinskii-Moriya interaction strength, $\hbar\omega_z$ the magnetic hard-axis anisotropy, and $\hbar\omega_{in}$ the magnetic in-plane anisotropy. Here, $\hat{\mathbf{z}}$ is the unit direction along the *z*-axis, and $\boldsymbol{\varphi}_0$ defines the in-plane easy axis direction, $\pm\boldsymbol{\varphi}_0$. For concreteness, we take $\boldsymbol{\varphi}_0 = (\sqrt{3}/2, 1/2, 0)$. We measured anisotropic magnetoresistance on a $Ni_{1/3}NbS_2$ device, indicating magnetic in-plane anisotropy (Fig. S2). This justifies the inclusion of the magnetic in-plane anisotropy in Eq. (1). The spin dynamics can be described by the Landau-Lifshitz-Gilbert equation,

$$\dot{\mathbf{S}}_i = -\mathbf{S}_i \times \mathbf{H}_i^{eff} + \alpha \mathbf{S}_i \times \dot{\mathbf{S}}_i + \mathbf{T}_i^{SOT}, \qquad (2)$$

where $\mathbf{H}_i^{eff} = -\partial H_{spin}/\hbar\partial \mathbf{S}_i$ is the effective magnetic field acting on layer $i$, $\alpha$ the Gilbert damping parameter, and $T_i^{\text{SOT}}$ the spin-orbit torque [35]. Due to the inversion symmetry breaking of Ni$_{1/3}$NbS$_2$, the spin-orbit torque is given by $T_i^{\text{SOT}} = \omega_{\text{FL}} S_i \times p + \omega_{\text{DL}} S_i \times (S_i \times p)$, where $\omega_{\text{FL}}$ ($\omega_{\text{DL}}$) is the field (antidamping)-like spin-orbit torque strength and $p \propto \hat{z} \times \hat{I}$ is spin polarization with $\hat{I}$ the charge current unit direction. The field (antidamping)-like spin-orbit torque gives rise to a staggered (uniform) torque [36]. This form of spin-orbit torque results from the broken global inversion symmetry in the bulk crystal structure [37] and spin-orbit coupling [14,36].

We define the local Néel order vector $n_i$ and the local ferromagnetic order vector $m_i$ as $n_i = (S_{2i} - S_{2i-1})/2$ and $m_i = (S_{2i} + S_{2i-1})/2$, respectively. Both $n_i$ and $m_i$ form helical textures. For the parameters in the Methods or Supplemental Note 1 [38], $|n_i| = 0.9989$ and $|m_i| = 0.04684$ for all $i$ in equilibrium. We also define the net Néel order $N_{\text{net}} = \sum_i^N (-1)^i S_i / N$, and the net magnetization $M_{\text{net}} = \sum_i^N S_i / N$. These two vectors are orthogonal to each other. Their magnitudes are $|N_{\text{net}}| = 2.051 \times 10^{-2}$ and $|M_{\text{net}}| = 9.619 \times 10^{-4}$ for $N=66$ in equilibrium. Here, the relation between the local order parameters and the net order parameters is $|n_i|/|m_i| \simeq |N_{\text{net}}|/|M_{\text{net}}|$. Because $|N_{\text{net}}| \gg |M_{\text{net}}|$, one may be inclined to analyze antiferromagnetic spin-orbit torque through Néel order dynamics. However, our simulation indicates that in the helical antiferromagnet, $M_{\text{net}}$ governs the antiferromagnetic control despite the smallness of $|M_{\text{net}}|$. This is a key difference between collinear antiferromagnet dynamics and helical antiferromagnet dynamics. As demonstrated in the End Matter section, $T_i^{\text{SOT}}$ rotates $M_{\text{net}}$ which acts as a lever to rotate the entire spin texture of the helical antiferromagnet.

**Current-controlled resistance change and its symmetry**

Figs. 2(h-i) show the optical image of our Ni$_{1/3}$NbS$_2$ nanoflake device with a thickness of ~50 nm and the schematic of the writing and reading functions. Based on our spin-orbit torque simulations above, we would like to elucidate the writing and reading principles for discussion here briefly.

Supposing one applies writing current $I_{\text{write1}}$ from left to right (black arrow), it will tend to rotate the weak ferromagnetic order toward a perpendicular direction by spin-orbit torque (black double-headed arrow). The entire helical order is rotated simultaneously, locked with the weak ferromagnetic order. When writing current $I_{\text{write2}}$ is applied from bottom to top (red arrow), the weak ferromagnetic order tends to align toward a horizontal direction (red double-headed arrow), accompanied by the helical order rotation. Interestingly, the current-induced modification of the helical spins by spin-orbit torque can be electrically read out through in-plane AMR (using longitudinal resistance $R_\parallel$) [39-41] and PHE (using transverse resistance $R_\perp$) [39-41]. Both effects exhibit the sinusoidal angular dependences of $2\theta$ with the mutual phase shift of 90º for $2\theta$, where $\theta$ is the reading angle between the reading current and the spin direction. Indeed, Fig. 2(j) demonstrates two distinct transverse resistance levels with a writing current of 5 mA, applied along $I_{\text{write1}}$ and $I_{\text{write2}}$ directions at reading angle $\theta=45º$. Similar current-driven behaviour has also been demonstrated in a much thinner device with a thickness of ~19 nm (Fig. S1).

To further validate the physical scenario experimentally, we rotate the reading geometry to check the symmetry of the current-controlled resistance change. Fig. 3(a) displays four cases with the reading angles $\theta=45º$, 90º, 135º, and 180º. In these four cases, the in-plane AMR and PHE should reach their minima and maxima successively, considering their $\cos 2\theta$ and $\sin 2\theta$ dependences. Here, rather than using the exact value of $R_\parallel$ and $R_\perp$, we focus on the resistance changes $\Delta R_\parallel$, $\Delta R_\perp$ at $I_{\text{write1}}$ and $I_{\text{write2}}$, where $\Delta R_\parallel = R_\parallel - \left(\left(R_\parallel(I_{\text{write1}}) + R_\parallel(I_{\text{write2}})\right)/2\right)$ and $\Delta R_\perp = R_\perp - \left(\left(R_\perp(I_{\text{write1}}) + R_\perp(I_{\text{write2}})\right)/2\right)$. It can capture the magnetic state's change and, thus, the antiferromagnetic spin-orbit torque. A more meaningful parameter is the relative resistance change $\Delta R_\parallel/R_\parallel$ or $\Delta R_\perp/R_\parallel$ as frequently used in the previous reports [5,14].

As shown in Fig. 3(b) for $\theta=45°$, the relative resistance change $\Delta R_\parallel/R_\parallel$ is nearly zero, while $\Delta R_\perp/R_\parallel$ increases in the negative (positive) direction as $I_{\text{write1}}$ ($I_{\text{write2}}$) increases with the difference illustrated in Fig. 3(f). For $\theta=45°$, the difference between the relative longitudinal resistance change $\left(\Delta R_\parallel(\text{red}) - \Delta R_\parallel(\text{black})\right)/R_\parallel$ is nearly zero, but the difference between the relative transverse resistance change $\left(\Delta R_\perp(\text{red}) - \Delta R_\perp(\text{black})\right)/R_\parallel$ is positive. Figs. 3(c-e) and 3(g-i) represent corresponding results for $\theta=90°$, 135°, and 180°, respectively. For $\theta=90°$, $\left(\Delta R_\parallel(\text{red}) - \Delta R_\parallel(\text{black})\right)/R_\parallel$ reaches a negative value, and $\left(\Delta R_\perp(\text{red}) - \Delta R_\perp(\text{black})\right)/R_\parallel$ turns to nearly zero. For $\theta=135°$, $\left(\Delta R_\parallel(\text{red}) - \Delta R_\parallel(\text{black})\right)/R_\parallel$ is nearly zero, and $\left(\Delta R_\perp(\text{red}) - \Delta R_\perp(\text{black})\right)/R_\parallel$ is negative. For $\theta=180°$, $\left(\Delta R_\parallel(\text{red}) - \Delta R_\parallel(\text{black})\right)/R_\parallel$ is positive, and $\left(\Delta R_\perp(\text{red}) - \Delta R_\perp(\text{black})\right)/R_\parallel$ is almost zero. All these symmetry observations support the antiferromagnetic spin-orbit torque scenario, i.e., the spin-orbit torque rotates the spin texture, and the spin-related in-plane AMR and PHE reach their minima and maxima with sign alternation successively following their $\cos 2\theta$ and $\sin 2\theta$ dependences.

We also inspect the current polarization dependency of the current-controlled resistance change. The resistance change is almost identical for +5 and -5 mA for both $I_{\text{write1}}$ and $I_{\text{write2}}$ [ Fig. S5(a)]. Moreover, its dependence on the magnitudes of the current $|I_{\text{write1}}|$ or $|I_{\text{write2}}|$ remains unchanged when the sign of the current polarization is reversed [Fig. S5(b)]. Our simulation result of $\delta \ll 1$ case [Figs. 2(f-g)], whose antidamping-like spin-orbit torque is dominant, can explain its independence to the writing current polarization. In addition, considering that the polarization reversal amounts to the shift of $\theta$ by $\pi$, such independence to the current polarization is consistent with the sinusoidal dependences of $2\theta$ for the in-plane AMR and PHE.

To summarize the symmetry result, we plot the relative resistance change $\Delta R_\perp/R_\parallel$ (red circle) and $\Delta R_\parallel/R_\parallel$ (black circle) as a function of the reading angle $\theta$ in Fig. 4(a). The red and black curves depict the expected $\cos 2\theta$ and $\sin 2\theta$ dependences of the AMR and the PHE, respectively.

**Temperature-current mapping of the current-controlled resistance change**
Next, we explore the temperature dependence of the current-controlled resistance change. For each reading angle $\theta=45°$, 90°, 135°, and 180°, either $\Delta R_\parallel/R_\parallel$ or $\Delta R_\perp/R_\parallel$ is nonzero. The nonvanishing quantity is probed as a function of temperature and writing-current amplitude. Fig. 4(b) demonstrates that for $\theta=45°$, the relative resistance change $\Delta R_\perp/R_\parallel$ is positive and features a transition-like decrease near ~85 K upon increasing temperature. It becomes clearer in the contour plot in Fig. 4(c). This feature is reminiscent of the paramagnetic-to-antiferromagnetic transition near $T_N$ of 85 K, pointing to a magnetism origin of the current-controlled resistance change. The spin-orbit torque scenario should account for the observed current-controlled resistance change.

In summary, we report the current control of the helical antiferromagnetic order using a vdW helical antiferromagnet. It is the first demonstration of the antiferromagnetic spin-orbit torque in a single helical antiferromagnet nanoflake device, combined with theory and experiment. Remarkably, the weak magnetic dipole (or net magnetization) of the helical antiferromagnetic order is strongly coupled to the entire helical antiferromagnetic order and acts as a lever to collectively allow the electrical control of the helical antiferromagnetic spin texture. We expect our results to help understand Néel order dynamics in the helical antiferromagnet.

**Acknowledgements**
We acknowledge Kyung-Jin Lee for his helpful comments on our work. The work at CQM and SNU was supported by the Samsung Science & Technology Foundation (Grant No. SSTF-BA2101-05). One of the authors (J.-G.P.) was partly funded by the Leading Researcher Program of the National Research Foundation of Korea (Grant Nos. 2020R1A3B2079375 and RS-2020-NR049405). The theoretical works at the POSTECH were funded by the National Research Foundation (NRF) of Korea (Grant Nos. 2020R1A2C2013484 and RS-2024-00410027). In addition, the Samsung Advanced Institute of Technology also supported this work at both SNU and POSTECH. This work was partly performed at the SIM beamline of the Swiss Light Source (SLS), Paul Scherrer Institut, Villigen, Switzerland. One of the authors (J.G.P.) acknowledges the hospitality of the Indian Institute of Science, where the manuscript was finalised, and the financial support of the Infosys Foundation.

**References**

[1] K. Olejnik, T. Seifert, Z. Kaspar, V. Novak, P. Wadley, R. P. Campion, M. Baumgartner, P. Gambardella, P. Nemec, J. Wunderlich, J. Sinova, P. Kuzel, M. Muller, T. Kampfrath, and T. Jungwirth, Terahertz electrical writing speed in an antiferromagnetic memory, Sci. Adv. **4**, eaar3566 (2018).

[2] T. Jungwirth, X. Marti, P. Wadley, and J. Wunderlich, Antiferromagnetic spintronics, Nat. Nanotechnol. **11**, 231 (2016).

[3] O. Gomonay, T. Jungwirth, and J. Sinova, Concepts of antiferromagnetic spintronics, Phys. Status Solidi Rapid Res. Lett. **11**, 1700022 (2017).

[4] V. Baltz, A. Manchon, M. Tsoi, T. Moriyama, T. Ono, and Y. Tserkovnyak, Antiferromagnetic spintronics, Rev. Mod. Phys. **90**, 015005 (2018).

[5] P. Wadley, B. Howells, J. Zelezny, C. Andrews, V. Hills, R. P. Campion, V. Novak, K. Olejnik, F. Maccherozzi, S. S. Dhesi, S. Y. Martin, T. Wagner, J. Wunderlich, F. Freimuth, Y. Mokrousov, J. Kunes, J. S. Chauhan, M. J. Grzybowski, A. W. Rushforth, K. W. Edmonds, B. L. Gallagher, and T. Jungwirth, Electrical switching of an antiferromagnet, Science **351**, 587 (2016).

[6] C. C. Chiang, S. Y. Huang, D. Qu, P. H. Wu, and C. L. Chien, Absence of Evidence of Electrical Switching of the Antiferromagnetic Néel Vector, Phys. Rev. Lett. **123**, 227203 (2019).

[7] H. Wu, H. Zhang, B. Wang, F. Gross, C. Y. Yang, G. Li, C. Guo, H. He, K. Wong, D. Wu, X. Han, C. H. Lai, J. Grafe, R. Cheng, and K. L. Wang, Current-induced Neel order switching facilitated by magnetic phase transition, Nat. Commun. **13**, 1629 (2022).

[8] J. Masell, X. Yu, N. Kanazawa, Y. Tokura, and N. Nagaosa, Combing the helical phase of chiral magnets with electric currents, Phys. Rev. B **102**, 180402(R) (2020).

[9] H. Masuda, T. Seki, J. I. Ohe, Y. Nii, H. Masuda, K. Takanashi, and Y. Onose, Room temperature chirality switching and detection in a helimagnetic $MnAu_2$ thin film, Nat. Commun. **15**, 1999 (2024).

[10] N. Jiang, Y. Nii, H. Arisawa, E. Saitoh, and Y. Onose, Electric current control of spin helicity in an itinerant helimagnet, Nat. Commun. **11**, 1601 (2020).

[11] T. Higo, K. Kondou, T. Nomoto, M. Shiga, S. Sakamoto, X. Chen, D. Nishio-Hamane, R. Arita, Y. Otani, S. Miwa, and S. Nakatsuji, Perpendicular full switching of chiral antiferromagnetic order by current, Nature **607**, 474 (2022).

[12] J. Y. Yoon, P. Zhang, C. T. Chou, Y. Takeuchi, T. Uchimura, J. T. Hou, J. Han, S. Kanai, H. Ohno, S. Fukami, and L. Liu, Handedness anomaly in a non-collinear antiferromagnet under spin-orbit torque, Nat. Mater. **22**, 1106 (2023).

[13] T. Xu, H. Bai, Y. Dong, L. Zhao, H.-A. Zhou, J. Zhang, X.-X. Zhang, and W. Jiang, Robust spin torque switching of noncollinear antiferromagnet $Mn_3Sn$, APL Mater. **11**, 071116 (2023).

[14] N. L. Nair, E. Maniv, C. John, S. Doyle, J. Orenstein, and J. G. Analytis, Electrical switching in a magnetically intercalated transition metal dichalcogenide, Nat. Mater. **19**, 153 (2020).

[15] E. Maniv, N. L. Nair, S. C. Haley, S. Doyle, C. John, S. Cabrini, A. Maniv, S. K. Ramakrishna, Y. L. Tang, P. Ercius, R. Ramesh, Y. Tserkovnyak, A. P. Reyes, and J. G. Analytis, Antiferromagnetic switching driven by the collective dynamics of a coexisting spin glass, Sci. Adv. **7**, eabd8452 (2021).

[16] K. S. Novoselov, A. Mishchenko, A. Carvalho, and A. H. Castro Neto, 2D materials and van der Waals heterostructures, Science **353**, aac9439 (2016).

[17] R. Bian, C. Li, Q. Liu, G. Cao, Q. Fu, P. Meng, J. Zhou, F. Liu, and Z. Liu, Recent progress in the synthesis of novel two-dimensional van der Waals materials, Natl. Sci. Rev. **9**, nwab164 (2022).

[18] J. Kim, S. Son, M. J. Coak, I. Hwang, Y. Lee, K. Zhang, and J.-G. Park, Observation of plateau-like magnetoresistance in twisted $Fe_3GeTe_2$/$Fe_3GeTe_2$ junction, J. Appl. Phys. **128**, 093901 (2020).

[19] J. G. Park, Opportunities and challenges of 2D magnetic van der Waals materials: magnetic graphene?, J. Phys. Condens. Matter **28**, 301001 (2016).

[20] J. U. Lee, S. Lee, J. H. Ryoo, S. Kang, T. Y. Kim, P. Kim, C. H. Park, J. G. Park, and H. Cheong, Ising-Type Magnetic Ordering in Atomically Thin $FePS_3$, Nano Lett. **16**, 7433 (2016).

[21] B. Huang, G. Clark, E. Navarro-Moratalla, D. R. Klein, R. Cheng, K. L. Seyler, D. Zhong, E. Schmidgall, M. A. McGuire, D. H. Cobden, W. Yao, D. Xiao, P. Jarillo-Herrero, and X. Xu, Layer-dependent ferromagnetism in a van der Waals crystal down to the monolayer limit, Nature **546**, 270 (2017).

[22] C. Gong, L. Li, Z. Li, H. Ji, A. Stern, Y. Xia, T. Cao, W. Bao, C. Wang, Y. Wang, Z. Q. Qiu, R. J. Cava, S. G. Louie, J. Xia, and X. Zhang, Discovery of intrinsic ferromagnetism in two-dimensional van der Waals crystals, Nature **546**, 265 (2017).

[23] K.-X. Zhang, G. Park, Y. Lee, B. H. Kim, and J.-G. Park, Magnetoelectric effect in van der Waals magnets, npj Quantum Mater. **10**, 6 (2025).

[24] K. S. Burch, D. Mandrus, and J. G. Park, Magnetism in two-dimensional van der Waals materials, Nature **563**, 47 (2018).

[25] S. Kang, K. Kim, B. H. Kim, J. Kim, K. I. Sim, J.-U. Lee, S. Lee, K. Park, S. Yun, T. Kim, A. Nag, A. Walters, M. Garcia-Fernandez, J. Li, L. Chapon, K.-J. Zhou, Y.-W. Son, J. H. Kim, H. Cheong, and J.-G. Park, Coherent many-body exciton in van der Waals antiferromagnet $NiPS_3$, Nature **583**, 785 (2020).

[26] K. S. Burch, Electric switching of magnetism in 2D, Nat. Nanotechnol. **13**, 532 (2018).

[27] X. Wang, J. Tang, X. Xia, C. He, J. Zhang, Y. Liu, C. Wan, C. Fang, C. Guo, W. Yang, Y. Guang, X. Zhang, H. Xu, J. Wei, M. Liao, X. Lu, J. Feng, X. Li, Y. Peng, H. Wei, R. Yang, D. Shi, X. Zhang, Z. Han, Z. Zhang, G. Zhang, G. Yu, and X. Han, Current-driven magnetization switching in a van der Waals ferromagnet $Fe_3GeTe_2$, Sci. Adv. **5**, eaaw8904 (2019).

[28] K. Zhang, S. Han, Y. Lee, M. J. Coak, J. Kim, I. Hwang, S. Son, J. Shin, M. Lim, D. Jo, K. Kim, D. Kim, H.-W. Lee, and J.-G. Park, Gigantic current control of coercive field and magnetic memory based on nm-thin ferromagnetic van der Waals $Fe_3GeTe_2$, Adv. Mater. **33**, 2004110 (2021).

[29] K. Zhang, Y. Lee, M. J. Coak, J. Kim, S. Son, I. Hwang, D. S. Ko, Y. Oh, I. Jeon, D. Kim, C. Zeng, H.-W. Lee, and J.-G. Park, Highly efficient nonvolatile magnetization switching and multi-level states by current in single van der Waals topological ferromagnet $Fe_3GeTe_2$, Adv. Funct. Mater. **31**, 2105992 (2021).

[30] K.-X. Zhang, H. Ju, H. Kim, J. Cui, J. Keum, J.-G. Park, and J. S. Lee, Broken inversion symmetry in van der Waals topological ferromagnetic metal iron germanium telluride, Adv. Mater. **36**, 2312824 (2024).

[31]  H. Wang, X. Y. Li, Y. Wen, R. Q. Cheng, L. Yin, C. S. Liu, Z. W. Li, and J. He, Two-dimensional ferromagnetic materials: From materials to devices, Appl. Phys. Lett. **121**, 220501 (2022).

[32]  J. Keum, K.-X. Zhang, S. Cheon, H. Kim, J. Cui, G. Park, Y. Chang, M. Kim, H.-W. Lee, and J.-G. Park, Novel magnetic-field-free switching behavior in vdW-magnet/oxide heterostructure, Adv. Mater., 2412037 (2025).

[33]  K. Lu, D. Sapkota, L. DeBeer-Schmitt, Y. Wu, H. B. Cao, N. Mannella, D. Mandrus, A. A. Aczel, and G. J. MacDougall, Canted antiferromagnetic order in the monoaxial chiral magnets $V_{1/3}TaS_2$ and $V_{1/3}NbS_2$, Phys. Rev. Mater. **4**, 054416 (2020).

[34]  Y. An, P. Park, C. Kim, K. Zhang, H. Kim, M. Avdeev, J. Kim, M.-J. Han, H.-J. Noh, S. Seong, J.-S. Kang, H.-D. Kim, and J.-G. Park, Bulk properties of the chiral metallic triangular antiferromagnets $Ni_{1/3}NbS_2$ and $Ni_{1/3}TaS_2$, Phys. Rev. B **108**, 054418 (2023).

[35]  A. Manchon, J. Železný, I. M. Miron, T. Jungwirth, J. Sinova, A. Thiaville, K. Garello, and P. Gambardella, Current-induced spin-orbit torques in ferromagnetic and antiferromagnetic systems, Rev. Mod. Phys. **91**, 035004 (2019).

[36]  J. Zelezny, H. Gao, K. Vyborny, J. Zemen, J. Masek, A. Manchon, J. Wunderlich, J. Sinova, and T. Jungwirth, Relativistic Neel-order fields induced by electrical current in antiferromagnets, Phys. Rev. Lett. **113**, 157201 (2014).

[37]  H. Kurebayashi, J. Sinova, D. Fang, A. C. Irvine, T. D. Skinner, J. Wunderlich, V. Novák, R. P. Campion, B. L. Gallagher, E. K. Vehstedt, L. P. Zârbo, K. Výborný, A. J. Ferguson, and T. Jungwirth, An antidamping spin–orbit torque originating from the Berry curvature, Nat. Nanotechnol. **9**, 211 (2014).

[38]  See Supplemental Materials for details on materials and methods, experiments, theoretical calculations, and additional experimental/simulated data and discussions, which includes Refs. [5,7,14,15,46-55].

[39]  A. A. Taskin, H. F. Legg, F. Yang, S. Sasaki, Y. Kanai, K. Matsumoto, A. Rosch, and Y. Ando, Planar Hall effect from the surface of topological insulators, Nat. Commun. **8**, 1340 (2017).

[40]  S. Nandy, G. Sharma, A. Taraphder, and S. Tewari, Chiral Anomaly as the Origin of the Planar Hall Effect in Weyl Semimetals, Phys. Rev. Lett. **119**, 176804 (2017).

[41]  A. A. Burkov, Giant planar Hall effect in topological metals, Phys. Rev. B **96**, 041110(R) (2017).

[42]  L. Liu, C. F. Pai, Y. Li, H. W. Tseng, D. C. Ralph, and R. A. Buhrman, Spin-torque switching with the giant spin Hall effect of tantalum, Science **336**, 555 (2012).

[43]  J. Cui, K.-X. Zhang, and J.-G. Park, All van der Waals Three-Terminal SOT-MRAM Realized by Topological Ferromagnet $Fe_3GeTe_2$, Adv. Electron. Mater. **10**, 2400041 (2024).

[44]  I. M. Miron, K. Garello, G. Gaudin, P. J. Zermatten, M. V. Costache, S. Auffret, S. Bandiera, B. Rodmacq, A. Schuhl, and P. Gambardella, Perpendicular switching of a single ferromagnetic layer induced by in-plane current injection, Nature **476**, 189 (2011).

[45]  L. Liu, O. J. Lee, T. J. Gudmundsen, D. C. Ralph, and R. A. Buhrman, Current-induced switching of perpendicularly magnetized magnetic layers using spin torque from the spin Hall effect, Phys. Rev. Lett. **109**, 096602 (2012).

[46] S. Seki, X. Yu, S. Ishiwata, and Y. Tokura, Observation of skyrmions in a multiferroic material, Science **336**, 198 (2012).

[47] N. D. Khanh, T. Nakajima, X. Yu, S. Gao, K. Shibata, M. Hirschberger, Y. Yamasaki, H. Sagayama, H. Nakao, L. Peng, K. Nakajima, R. Takagi, T. H. Arima, Y. Tokura, and S. Seki, Nanometric square skyrmion lattice in a centrosymmetric tetragonal magnet, Nat. Nanotechnol. **15**, 444 (2020).

[48] H. Yoshimochi, R. Takagi, J. Ju, N. D. Khanh, H. Saito, H. Sagayama, H. Nakao, S. Itoh, Y. Tokura, T. Arima, S. Hayami, T. Nakajima, and S. Seki, Multistep topological transitions among meron and skyrmion crystals in a centrosymmetric magnet, Nat. Phys. **20**, 1001 (2024).

[49] S. S. Parkin, M. Hayashi, and L. Thomas, Magnetic domain-wall racetrack memory, Science **320**, 190 (2008).

[50] L. Thomas, R. Moriya, C. Rettner, and S. S. P. Parkin, Dynamics of Magnetic Domain Walls Under Their Own Inertia, Science **330**, 1810 (2010).

[51] S. Parkin and S. H. Yang, Memory on the racetrack, Nat. Nanotechnol. **10**, 195 (2015).

[52] Z. Luo, A. Hrabec, T. P. Dao, G. Sala, S. Finizio, J. Feng, S. Mayr, J. Raabe, P. Gambardella, and L. J. Heyderman, Current-driven magnetic domain-wall logic, Nature **579**, 214 (2020).

[53] M. Wu, T. Chen, T. Nomoto, Y. Tserkovnyak, H. Isshiki, Y. Nakatani, T. Higo, T. Tomita, K. Kondou, R. Arita, S. Nakatsuji, and Y. Otani, Current-driven fast magnetic octupole domain-wall motion in noncollinear antiferromagnets, Nat. Commun. **15**, 4305 (2024).

[54] C. O. Avci, K. Garello, M. Gabureac, A. Ghosh, A. Fuhrer, S. F. Alvarado, and P. Gambardella, Interplay of spin-orbit torque and thermoelectric effects in ferromagnet/normal-metal bilayers, Phys. Rev. B **90**, 224427 (2014).

[55] J. Godinho, H. Reichlova, D. Kriegner, V. Novak, K. Olejnik, Z. Kaspar, Z. Soban, P. Wadley, R. P. Campion, R. M. Otxoa, P. E. Roy, J. Zelezny, T. Jungwirth, and J. Wunderlich, Electrically induced and detected Neel vector reversal in a collinear antiferromagnet, Nat. Commun. **9**, 4686 (2018).

**End Matter**
**Detailed theoretical calculations and predictions**
We investigate the spin texture in the helical antiferromagnet $Ni_{1/3}NbS_2$ with one helical period ($N=66$) using the Landau-Lifshitz-Gilbert equation [Eq. (2)]. Fig. 2(a) shows the equilibrium helical spin texture of $S_i$. Let us now turn on the writing current $I_{write1}$ or $I_{write2}$, whose directions are along the $\hat{x}$ and $\hat{y}$ directions [[100] and [120], respectively, in Figs. 2 (h-i)]. As shown in Fig. 2(c), the steady-state helical antiferromagnetic spin texture in the presence of either $I_{write1}$ or $I_{write2}$ is rotated compared to the initial helical texture in Fig. 2(a). The spin-orbit torques do not significantly deform the helical spin texture except for its collective rotation around the $c$-axis. $M_{net}$ is rotated by the same amount by $I_{write1}$ or $I_{write2}$. Therefore, $M_{net}$ and helical antiferromagnetic spin texture are strongly connected with each other, and the net nonzero $M_{net}$ can be an easy way to understand the collective helical antiferromagnet dynamics caused by the spin-orbit torques. Interestingly, $M_{net}$ prefers to be directed in the $\hat{z} \times I_{write1(2)}$ by the spin-orbit torque. This behaviour is similar to the spin-orbit torque-induced rotation of the ferromagnetism in conventional heavy-metal/ferromagnet bilayer systems [37,42].

Figs. 2(d-e) show the azimuthal angle $\phi_{M_{net}}$ (the polar angle $\theta_{M_{net}}$) of $M_{net}$ at the steady state in the presence of the writing current along the $x$ or $y$ direction. Interestingly, the helical antiferromagnetic spin texture is rotated while the current is on. This rotation angle is almost equal to the current-induced change of $\phi_{M_{net}}$, implying a close link between the $M_{net}$ dynamics and the dynamics of the helical antiferromagnetic spin texture. To examine the possibility further, we investigate a one-dimensional spin chain with periodic boundary conditions (see Supplemental Note 2 [38]). In this case, we find that $M_{net}$ becomes essentially zero in clear contrast to the finite $M_{net}$ of a one-dimensional spin chain with free ends. In this special situation of vanishing $M_{net}$, we find that applying an in-plane current rotates the local spin directions by extremely tiny angles smaller than our numerical calculation accuracies. We thus conclude that the spin-orbit torque cannot rotate the helical antiferromagnetic order at all when $M_{net}$ vanishes. This result supports our interpretation that $M_{net}$ is important for the rotation of the helical antiferromagnetic order.

We examine which component of the spin-orbit torque is mainly responsible for the response of the helical antiferromagnetic order. For this, we vary the magnitudes of its field-like and antidamping-like components with the ratio ($\delta$) of the former component to the latter ranging from 0 to $\infty$. When the antidamping-like spin-orbit torque is dominant ($\delta \ll 1$), the current tilts $M_{net}$ by changing $\theta_{M_{net}}$ linearly in Fig. 2(e) and $\phi_{M_{net}}$ quadratically in Fig. 2(d), regardless of whether the current is applied along the $\pm x$ or $\pm y$ direction. When the field-like spin-orbit torque is dominant ($\delta \gg 1$), the current changes $\phi_{M_{net}}$ linearly [results for $\delta = 10$ and $\infty$ in Figs. 2(f-g)]. Electrical detections [anisotropic magnetoresistance (AMR) and planar Hall effect (PHE)] arise from the deviation of the azimuthal angle within the helical antiferromagnet. Thus, we can expect that for $\delta \ll 1$, the resistance remains independent of the writing current polarization, while the resistance for $\delta \gg 1$ depends on the writing current polarization. In the next section, we find that in the helical antiferromagnet $Ni_{1/3}NbS_2$, antidamping-like spin-orbit torque is a dominant contribution (i.e., $\delta \ll 1$).

Although the field-like spin-orbit torque is negligible in $Ni_{1/3}NbS_2$, its effect is still worth discussing since it illustrates an interesting difference between helical and collinear antiferromagnets. In collinear antiferromagnets, the field-like spin-orbit torque is irrelevant to the magnetization dynamics if the torque is staggered [36]. In contrast, the staggered field-like spin-orbit torque is relevant in helical antiferromagnets and can even control the order. This feature is similar to ferromagnets, for which both field-like and antidamping-like spin-orbit torque can achieve the spin reorientation, and is consistent with the importance of $M_{\text{net}}$ for the helical antiferromagnetic order dynamics.

We have analyzed the current-induced dynamics of 66 local spins using the Landau-Lifshitz-Gilbert equation. Our calculation treats all 66 spins as independent degrees of freedom. We have confirmed that the helical spin configuration in the helical antiferromagnet $Ni_{1/3}NbS_2$ is highly rigid, and the most relevant degree of freedom is the collective rotation of the entire spin configuration. Our theoretical analysis shows that the helical antiferromagnetic order can be electrically controlled through the spin-orbit torque. Its main effect is to induce the collective rotation of the helical antiferromagnetic order, which can well be represented by the magnetic dipole ($M_{\text{net}}$) dynamics associated with the helical antiferromagnetic spin order through the Dzyaloshinskii-Moriya interaction.

**Comparisons of critical current density in different systems**

The critical threshold tuning current is 0.5 mA, corresponding to a current density of $10^6$ A/cm$^2$ for $Ni_{1/3}NbS_2$ nanodevices. Regarding ferromagnetic systems, our critical current density (~$10^6$ A/cm$^2$) is similar to other van-der-Waals ferromagnetic systems like $Fe_3GeTe_2$ [28,29,43] (~$10^6$ A/cm$^2$) but orders of magnitude lower than conventional ferromagnetic spin-orbit torque systems like FM/Pt [44,45] bilayer (~$10^8$ A/cm$^2$). Regarding antiferromagnetic systems, our critical current density is above one order higher than $Fe_{1/3}NbS_2$ [14] (~$10^4$ A/cm$^2$) but similar to the famous CuMnAs [5] system (~$10^6$ A/cm$^2$).

**Additional discussions on switching's temperature dependence and symmetry, the low possibility of the current-driven ferromagnetic domain wall, and the critical novel differences in our work.**

As a passing remark, there is a bump around 40 K, which makes the temperature dependence nonmonotonic. We suspect it might be due to subtle magnetic structure changes or spin fluctuations, the origin of which would require high-resolution neutron diffraction studies under the field and current and is an interesting subject of future works. Figs. S6, S7, and S8 show similar transition-like characters for $\theta$=90º, 135º, and 180º, but with negative $\Delta R_\parallel / R_\parallel$, negative $\Delta R_\perp / R_\parallel$, and positive $\Delta R_\parallel / R_\parallel$, respectively, consistent with the symmetry results in Figs. 3 and 4(a). Note that we have discussed the low possibility of the current-driven ferromagnetic domain wall as an account for our work in Supplemental Note 4 [38]. In addition, we have made a clear comparison to explicitly clarify the four critical differences between our work and the previous spin-glass-states-dominated case [14,15] in Supplemental Note 5 [38].

**Figures**

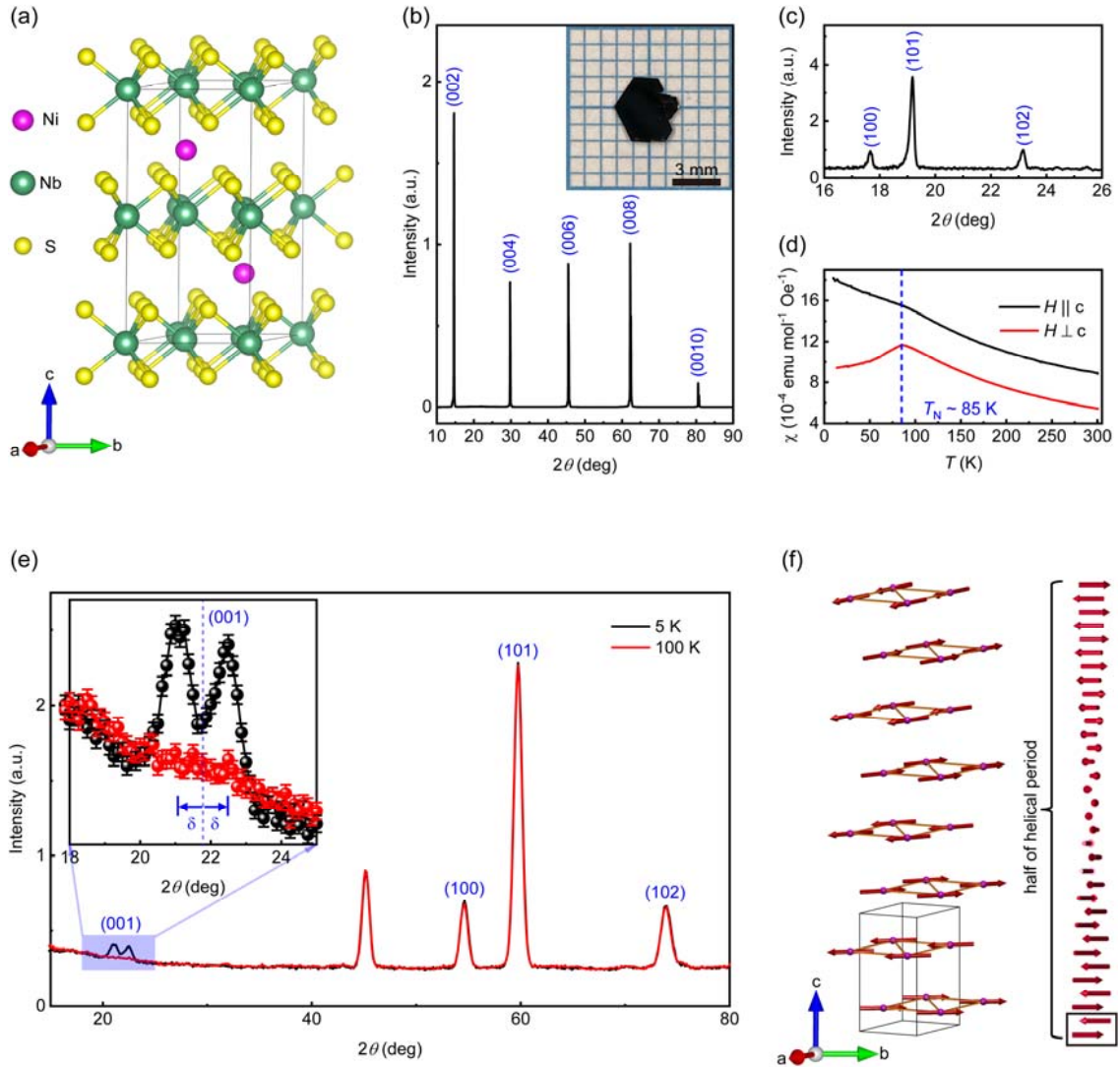

Fig. 1. (a) Structure of Ni$_{1/3}$NbS$_2$. (b) XRD pattern and optical image of a typical single crystal. (c) Powder XRD pattern. (d) Susceptibility-temperature curves with the Néel temperature $T_N$ of 85 K. (e) Neutron powder diffraction pattern at 5 and 100 K with a wavelength of 0.474 nm. The inset shows the magnetic satellite peaks (0, 0, 1±$\delta$) near the (0,0,1) position at 5 K below $T_N$, indicating a helical spin texture with a periodicity of 66 layers. (f) Helical in-plane spin texture of A-type model for antiferromagnet Ni$_{1/3}$NbS$_2$.

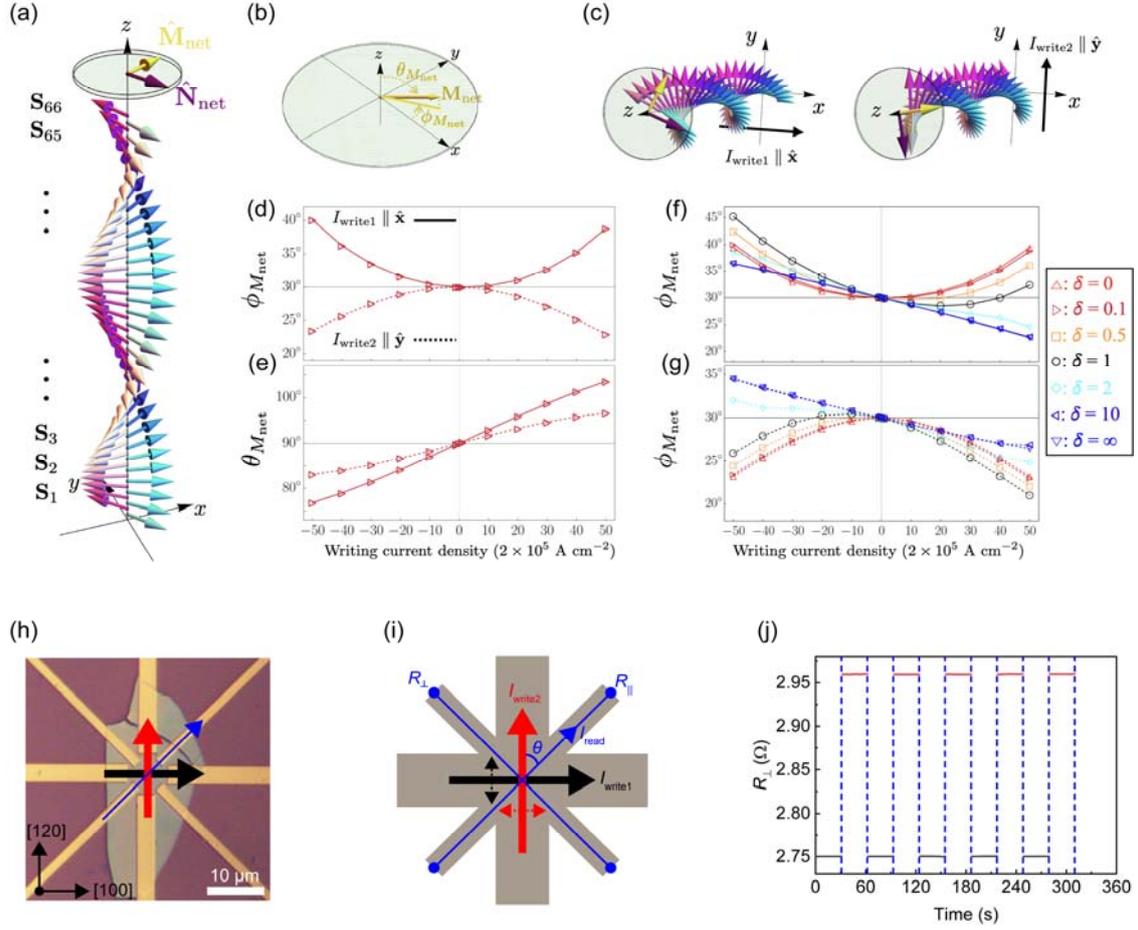

Fig. 2. Theoretical simulations of electrically tuning the helical spin texture. (a) The equilibrium state of helical antiferromagnetic spin texture without writing currents. (b) Schematic illustration of the net magnetization $\mathbf{M}_{net}$ within a spherical coordinate system with azimuthal angle $\phi_{M_{net}}$ and polar angle $\theta_{M_{net}}$. (c) Helical antiferromagnetic spin texture with the spin-orbit torques in the presence of writing current $I_{write1} \parallel x$ (left) and $I_{write2} \parallel y$ (right). (d-e) $\phi_{M_{net}}$ and $\theta_{M_{net}}$ versus writing current density at $\delta = 0.1$, where $\delta$ is the field-like and antidamping-like spin-orbit torque ratio. (f-g) $\phi_{M_{net}}$ versus writing current density at various $\delta$ for $I_{write1}$ (f) and $I_{write2}$ (g). When $\omega_{FL} < \omega_{DL}$ (red-colored symbols), $\phi_{M_{net}}$ exhibits symmetry. Conversely, for $\omega_{FL} > \omega_{DL}$ (blue-coloured symbols), $\phi_{M_{net}}$ shows asymmetry. (h-i) Optical image of a Ni$_{1/3}$NbS$_2$ nanoflake device (h) and the corresponding measurement schematic (i). The writing current is applied as $I_{write1}$ or $I_{write2}$. The reading current is applied with a reading angle $\theta$ to the writing current $I_{write2}$, while monitoring the longitudinal resistance $R_\parallel$ and transverse resistance $R_\perp$. The black (red) double-headed arrow represents the expected preferred spin direction by spin-orbit torque while applying the writing current $I_{write1}$ ($I_{write2}$). (j) Typical current controlled resistance change, e.g., the $R_\perp$ jumps between two distinct resistance levels for $I_{wirte1}$ (black) and $I_{wirte2}$ (red) under 45 K at a writing current of 5 mA and $\theta=45°$.

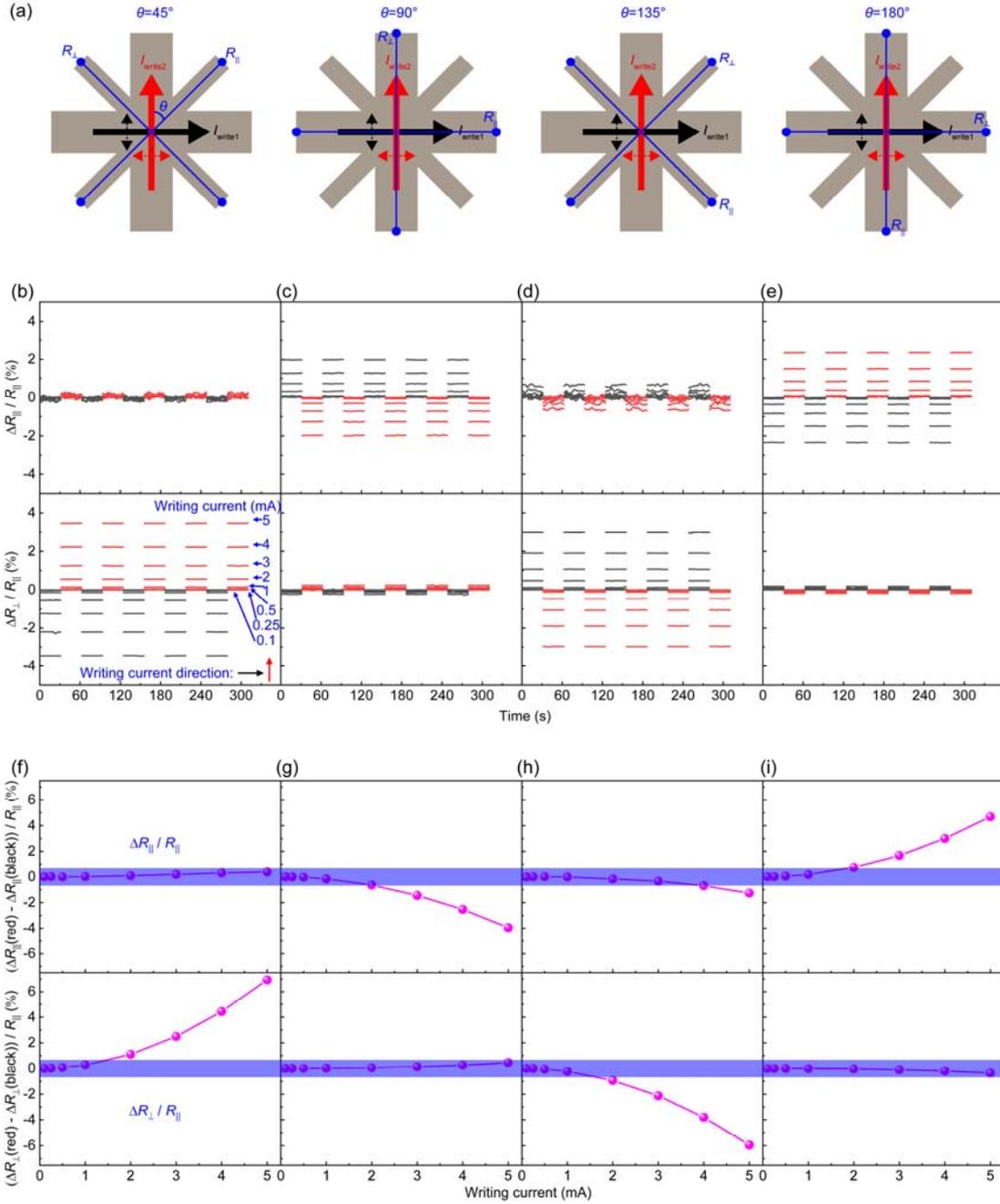

Fig. 3. Symmetry of the current-controlled resistance change. (a) Four cases with $\theta$=45°, 90°, 135°, and 180°, respectively. (b-e) Relative resistance change $\Delta R_\parallel/R_\parallel$ and $\Delta R_\perp/R_\parallel$ for $\theta$=45° (b), 90° (c), 135° (d), 180°(e), respectively. The black and red curves indicate the relative resistance change for each $I_{write1}$ and $I_{write2}$. (f-i) The difference of relative resistance change, *i.e.*, $\left(\Delta R_\parallel(\text{red}) - \Delta R_\parallel(\text{black})\right)/R_\parallel$ and $\left(\Delta R_\perp(\text{red}) - \Delta R_\perp(\text{black})\right)/R_\parallel$ versus writing-current amplitude for $\theta$=45° (f), 90° (g), 135° (h), and 180° (i), respectively. The blue bars shadow the region of nearly zero difference.

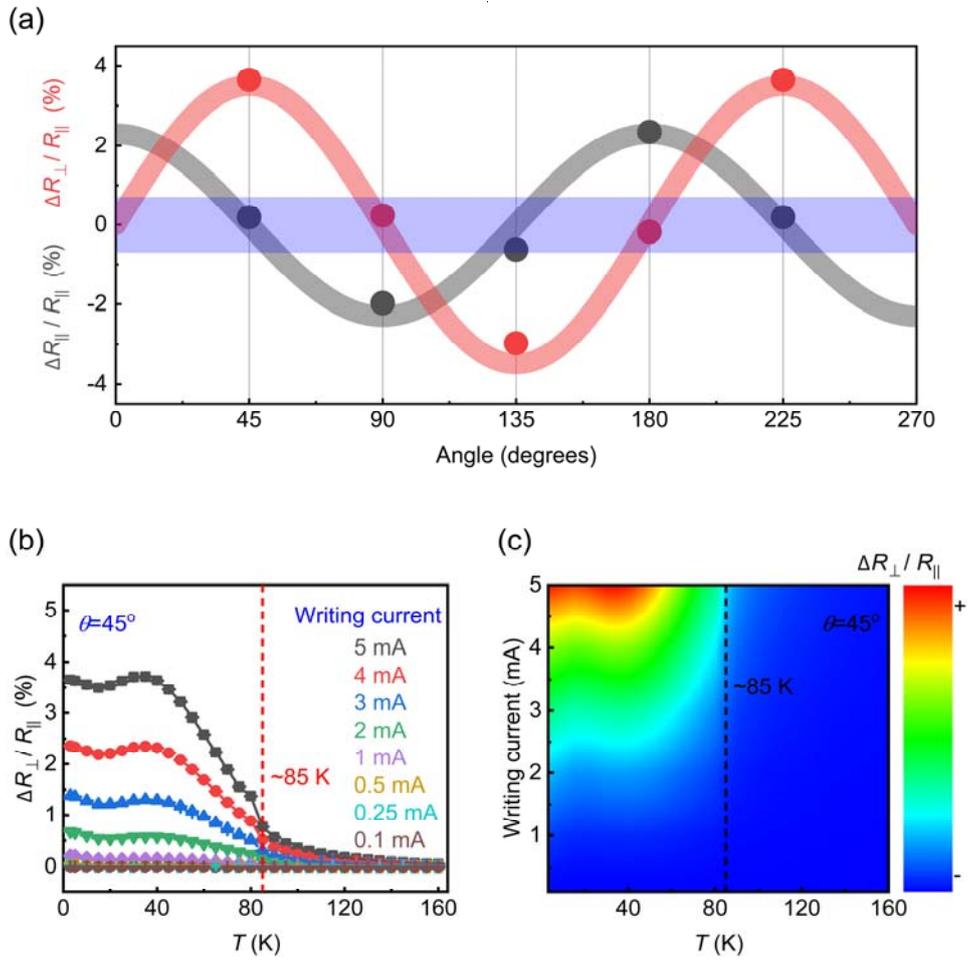

Fig. 4. Reading angle and temperature dependence. (a) Sinusoidal $\Delta R_\perp/R_\parallel$-$\theta$ and $\Delta R_\parallel/R_\parallel$-$\theta$ curves. (b) $\Delta R_\perp/R_\parallel$-$T$ curves for $\theta$=45° at various writing-current amplitudes. (c) Temperature-current mapping of $\Delta R_\perp/R_\parallel$ for $\theta$=45°.

# Supplemental Material for:

# Current-driven collective control of helical spin texture in van der Waals antiferromagnet

Kai-Xuan Zhang,[1,2†*] Suik Cheon,[3†] Hyuncheol Kim,[1,2†] Pyeongjae Park,[1,2¶] Yeochan An,[1,2] Suhan Son,[1,2‡] Jingyuan Cui,[1,2§] Jihoon Keum,[1,2] Joonyoung Choi,[4] Younjung Jo,[4] Hwiin Ju,[5] Jong-Seok Lee,[5] Youjin Lee,[1,2#] Maxim Avdeev,[6,7] Armin Kleibert,[8] Hyun-Woo Lee,[3,9*] Je-Geun Park[1,2*]

[†] These authors contributed equally to this work.
[*] Corresponding authors: Kai-Xuan Zhang (email: kxzhang.research@gmail.com), Hyun-Woo Lee (email: hwl@postech.ac.kr), and Je-Geun Park (email: jgpark10@snu.ac.kr).

This PDF file includes:
    Materials and Methods
    Supplemental Notes
    Supplemental Figures S1 to S12
    Supplemental References

**Materials and Methods**

Preparation of $Ni_{1/3}NbS_2$ single crystals and nanoflake devices

High-quality $Ni_{1/3}NbS_2$ single crystals were grown via a two-step process. First, the solid-state reaction pre-made the polycrystalline $Ni_{1/3}NbS_2$ to make the sample chemically homogeneous. High-purity element (Ni, Nb, S) powders were mixed in a ratio of 1.1:3:6 and sealed in a vacuum quartz tube, kept in a furnace at 900°C for sintering for one week. Then, the single crystals were grown from those prepared polycrystals by a chemical vapour transport method, using iodine as the transport agent. The tube was placed in a two-zone furnace to grow single crystals with a temperature gradient of 940 °C (source) to 860 °C (sink) for ten days.

$Ni_{1/3}NbS_2$ nanoflakes were exfoliated from the as-grown single crystals by a mechanical exfoliation method with Scotch tape. The nanometers-thin nanoflakes were exfoliated inside the glove box filled with Argon gas, although $Ni_{1/3}NbS_2$ is air-stable. Before removal, the samples were put into a load-lock box inside the glove box. Then, the Polymethyl methacrylate (PMMA) polymer was spin-coated onto $Ni_{1/3}NbS_2$ nanoflakes and then baked at 130 °C for 1.5 min for the following standard electron beam lithography (EBL). After EBL, 100/10 nm Au/Ti electrodes were evaporated onto the nanodevice by electron beam evaporation under a high vacuum (<$10^{-5}$ Pa).

Electrical transport measurements

The transport measurements were performed using a resistivity probe operated inside a cryostat down to 2 K. The writing current was applied by Keithley 6220, while the reading resistance was monitored using a standard lock-in technique by Stanford SR830. Gold wires were wire bonded to connect the electronic chip to the sample's pad electrodes. An antistatic wrist strap was used during the operation to prevent the possible damage of electrostatic discharges or shocks to the sample. In-plane anisotropic magnetoresistance measurements were carried out using a rotation probe, with an in-plane magnetic field up to 9 T rotating in the sample plane.

Theoretical analysis and spin-orbit torque simulations

The Landau-Lifshitz-Gilbert equations were numerically solved by using Mathematica. We set realistic values for the coefficient used in the calculation. The antiferromagnetic spin exchange interaction is $\hbar\omega_{\mathrm{ex}} = 3.5408 \text{ meV}$, the Dzyalonshinskii-Moriya interaction is $\hbar\omega_{\mathrm{DMI}} = 0.094\,\hbar\omega_{\mathrm{ex}}$, and the phenomenological magnetic hard axis anisotropy $\hbar\omega_{\mathrm{z}} = 1.4658 \text{ μeV}$ [see Supplemental Note 1]. Also, we assume that the phenomenological magnetic in-plane anisotropy $\hbar\omega_{\mathrm{in}} = 0.2\,\hbar\omega_{\mathrm{z}}$. The spin-orbit torque ratio between $\omega_{\mathrm{FL}}$ and $\omega_{\mathrm{DL}}$ is $\delta$. When $\delta < 1$, $\omega_{\mathrm{DL}} = \omega_{\mathrm{SOT}}$ and $\omega_{\mathrm{FL}} = \delta\omega_{\mathrm{SOT}}$. When $\delta > 1$, $\omega_{\mathrm{DL}} = \omega_{\mathrm{SOT}}/\delta$ and $\omega_{\mathrm{FL}} = \omega_{\mathrm{SOT}}$. Here, $\omega_{\mathrm{SOT}} = \gamma_e \frac{\hbar\theta_{\mathrm{SH}}}{2e}\frac{J_e A}{\mu_s}$, where $\gamma_e$ the gyromagnetic ratio, $\theta_{\mathrm{SH}} = 0.01$ the spin Hall angle, $J_e$ the charge current density, $\mu_s$ the local atomic moment, and $A$ the unit cell area of the single magnetic layer. In addition, we assume the Gilbert damping parameter to be 0.05.

**Supplemental Notes**

Note 1. Model Hamiltonian for $Ni_{1/3}NbS_2$

The helical magnetic ground state of $Ni_{1/3}NbS_2$ can be understood based on the following model Hamiltonian:

$$H_{spin} = J_1 \sum_{i,j} \mathbf{S}_i \cdot \mathbf{S}_j + J_c \sum_{i,j} \mathbf{S}_i \cdot \mathbf{S}_j + \sum_{i,j} \mathbf{D}_c \cdot (\mathbf{S}_i \times \mathbf{S}_j) + K \sum_i (S_i^z)^2. \quad (S1)$$

The first two terms are ferromagnetic intra-layer ($J_1 < 0$) and antiferromagnetic inter-layer ($J_c > 0$) nearest-neighbour (NN) interactions. The third term denotes the Dzyaloshinskii–Moriya interaction between inter-layer NNs. $K$ ($> 0$) is phenomenological magnetic hard axis anisotropy to realize the spin configuration perpendicular to the $c$-axis.

First, the pitch of the helical order determines the ratio between $J_c$ and $|\mathbf{D}_c|$:

$$|\mathbf{D}_c|/J_c = \tan(\Delta\phi), \quad (S2)$$

where $\Delta\phi$ is the angle between the two magnetic moments connected by $J_c$. The neutron diffraction result of $Ni_{1/3}NbS_2$ gives $\Delta\phi \sim 5.4°$, and therefore $|\mathbf{D}_c|/J_c = 0.094$. The Curie-Weiss temperature of $Ni_{1/3}NbS_2$ fitted from the $M(T)$ data further yields a constraint of exchange parameters:

$$\theta_{CW} = -\frac{S(S+1)}{3k_B}\sum_i z_i J_i = -\frac{2S(S+1)}{k_B}(J_1 + J_c). \quad (S3)$$

Using $\theta_{CW} \sim -100$ K of $Ni_{1/3}NbS_2$ and assuming $S = 1$ as in the case of $Ni^{2+}$, Eq. S3 gives $J_1 + J_c = 2.075$ meV.

The two constraints mentioned above leave $J_1$ as the only parameter not determined, *i.e.*, $J_c$ and $|\mathbf{D}_c|$ are automatically determined from $J_1$. A reasonable estimation of $J_1$ was done by reproducing the transition temperature of $Ni_{1/3}NbS_2$ ($T_N = 85$ K) in classical Monte-Carlo (MC) simulation. We employed the Metropolis-Hastings algorithm combined with the simulated annealing method for the simulation, using an $8 \times 8 \times 66$ supercell with periodic boundary conditions. After iterating 6,000 MC steps for equilibration, 180,000

MC steps were used to calculate the heat capacity of the spin system. As a result, we obtained $J_1$ = -1.4658 meV and $J_c$ = 3.5408 meV. We estimated the strength of the Dzyaloshinskii–Moriya interaction and exchange and their relative ratio ($|\boldsymbol{D}_c|/J_c$ = 0.094) using the ground state within our model Hamiltonian, based on our experimental results and MC simulations. The phenomenological magnetic hard axis anisotropy was assumed to be the common value, e.g., $K$ = 0.1%$|J_1|$ = 0.001466 meV, which does not affect the ground state or the helical AFM texture in our spin-orbit torque simulations. Note that the simulated annealing down to 0.5 K yields the correct magnetic ground state.

Note 2. Periodic boundary conditions

In the main text, we solved the coupled Landau-Lifshitz-Gilbert equations for local spins with $N$ = 66. Our result demonstrates that it connects the two quantities: $\boldsymbol{M}_{\text{net}}$ and $\boldsymbol{N}_{\text{net}}$. In this supplementary note, we investigate whether $\boldsymbol{M}_{\text{net}}$ and $\boldsymbol{N}_{\text{net}}$ are strongly coupled even in a one-dimensional spin chain model with periodic conditions, not a one-dimensional spin chain model with free ends. Here, we note that under the periodic boundary condition, the two-fold in-plane magnetic anisotropy does not give rise to a specific direction of the equilibrium $\boldsymbol{M}_{\text{net}}$ and $\boldsymbol{N}_{\text{net}}$. Therefore, in this supplementary note, we neglect the in-plane magnetic anisotropy.

In the one-dimensional spin chain model, the periodic boundary condition couples the spin moments at the top and the bottom layers. For example, the exchange interaction in Eq. (1) of the model can be written as follows:

$$H_{\text{spin, ex}} = \hbar\omega_{\text{ex}}(\boldsymbol{S}_{66} \cdot \boldsymbol{S}_1 + \boldsymbol{S}_1 \cdot \boldsymbol{S}_2 + \cdots + \boldsymbol{S}_{65} \cdot \boldsymbol{S}_{66}). \tag{S4}$$

After numerical calculation, one obtains $|\boldsymbol{N}_{\text{net}}| \propto 10^{-8}, |\boldsymbol{M}_{\text{net}}| \propto 10^{-9}$, which are

significantly smaller than the value ($|N_{\text{net}}| = 2.051 \times 10^{-2}, |M_{\text{net}}| = 9.619 \times 10^{-4}$) for the case with free ends (that is, without the antiferromagnetic exchange coupling term $\hbar\omega_{\text{ex}}S_{66} \cdot S_1$).

Our investigation confirms that in the periodic boundary conditions, $M_{\text{net}}$ is almost negligible. Also, we cannot control the helical spin texture by using the spin-orbit torque. Thus, this result shows that $M_{\text{net}}$, strongly intertwined with the helical texture, plays a crucial role in manipulating the helical antiferromagnet. In addition, the helical antiferromagnet with the periodic boundary condition is an idealized scenario, and actual samples do not satisfy this boundary condition. Consequently, due to the finite size of the sample, the collective spins exhibit mostly in-plane $M_{\text{net}}$ and $N_{\text{net}}$, which are orthogonal to each other. Such a case allows us to manipulate the helical antiferromagnetic order using spin-orbit torques.

Note 3. Issues on Joule heating effect.

Unfortunately, Joule heating is unavoidable for any current-dependent measurement, especially for current-driven torque experiments with large currents. Nevertheless, the agreement between our experimental results and simulation results of the current-induced torque and the temperature-dependent transition-like feature reinforce our claim that the antiferromagnetic spin-orbit torque is the primary origin of the measurement results. However, one cannot entirely rule out the possibility that Joule heating may facilitate this process through thermal activation [1].

Note 4. Discussions on low-possibility current-driven ferromagnetic domain wall.

We want to discuss the low possibility of the current-driven ferromagnetic domain wall as an account for our work for the following reasons.

Current-driven ferromagnetic domain wall motion can cause resistance changes. However, the resistance change due to this mechanism is inconsistent with our experimental result for the following reasons. Suppose the current-driven ferromagnetic domain wall motion was the main reason for the resistance changes. In that case, the value of $R_\perp$ is determined by whether or not the domain wall passes the cross-junction of the eight-leg geometry [Fig. 2(i)]. Then, the measured value of $R_\perp$ should be highly stochastic since our experiment has no control over the domain wall's initial location. However, our measurements in Figs. 3 and 4 indicate that $R_\perp$ is not stochastic. Furthermore, the discrepancy persists even if the stochasticity is suppressed for unknown reasons, although it is unlikely. For instance, if the current-driven ferromagnetic domain wall motion were the main reason for the resistance changes, the value of $R_\perp$ as a function of current should increase abruptly since $R_\perp$ should change from one value to another value suddenly as the domain wall passes the cross-junction. In contrast, the measured data show a smooth increase with the current. Based on these observations, we conclude that the current-induced resistance change is not due to the current-driven ferromagnetic domain wall motion.

Moreover, please note that the critical switching current is 0.5 mA, corresponding to a low current density of $10^6$ A/cm$^2$. Such low current density generally cannot induce sizable ferromagnetic domain walls according to many nice Skyrmion and domain papers [2-9], most of which require orders of magnitude larger current density. We believe the low current density would not easily produce the ferromagnetic domain in the robust antiferromagnetic ground states.

In summary, our experimental and theoretical works have demonstrated that the antiferromagnetic SOT is the main reason for the observations, which otherwise cannot be strongly supported by the low-possibility current-driven ferromagnetic domain wall.

Note 5. Differences between our work and previous spin-glass-states-dominated case.

We want to emphasize the critical differences between our work and the previous Fe-intercalated-NbS$_2$ work [10]. First, Fe$_{1/3}$NbS$_2$ is an antiferromagnet with the collinear out-of-plane Néel order [10]. However, what is relevant for the electrical control is not the out-of-plane order but a tiny in-plane order [10] that coexists with the primary out-of-plane order. It was suggested [11] that the current-controlled resistance change in this material probably amounts to reorienting the tiny in-plane order (rather than the out-of-plane order). The in-plane order appears to be closely linked with the spin-glass order formation [11], but the nature and the dynamics of the tiny in-plane order are not clearly understood.

In contrast, the magnetic order of Ni$_{1/3}$NbS$_2$ is clearly characterized by the helical in-plane Néel order of A-type. Moreover, our experimental result can be understood more simply by our conclusion: the collective rotation of the helical order as the net magnetic moment, which arises from the Dzyaloshinskii-Moriya interaction of the system, is rotated by an in-plane current. Therefore, our work constitutes the first experimental demonstration of the helical order's electrical control and provides a simple picture to understand it.

Second, Fe-intercalated-NbS$_2$'s spin-glass states are deeply linked to this material's antiferromagnetic spin-orbit torque [11], and the involvement of the spin-glass order limits its applicability to specific low-temperature regimes. Meanwhile, its switching is sharply suppressed upon applying an external magnetic field due to the sensitivity to the magnetic field of the spin-glass states. By contrast, the spin order and the antiferromagnetic switching are stable against a magnetic field in the pure helical model $Ni_{1/3}NbS_2$. Despite the stability against a magnetic field, the helical spin texture can be effectively tuned at the whole temperature range below $T_{Néel}$ by an electrical current, as demonstrated in our work. Finally, it highlights the possibility of using the spin-orbit torque switching of helical antiferromagnets for reliable work in a complex electromagnetic environment.

Third, the antiferromagnetic spin-orbit torque of $Fe_{1/3}NbS_2$ was demonstrated only for thick films with 1~10 μm thickness [10,11], which are too thick for 2D-material-based spintronics and nanodevices. In comparison, we achieve the collective rotation of the entire helical spin texture by antiferromagnetic spin-orbit torque in ~19 nm thick devices, over 100 times thinner than previous devices. Our work provides the realization of antiferromagnetic spin-orbit torque in a single vdW nanoflake device down to 19 nm, which is the record low thickness.

**Supplemental Figures**

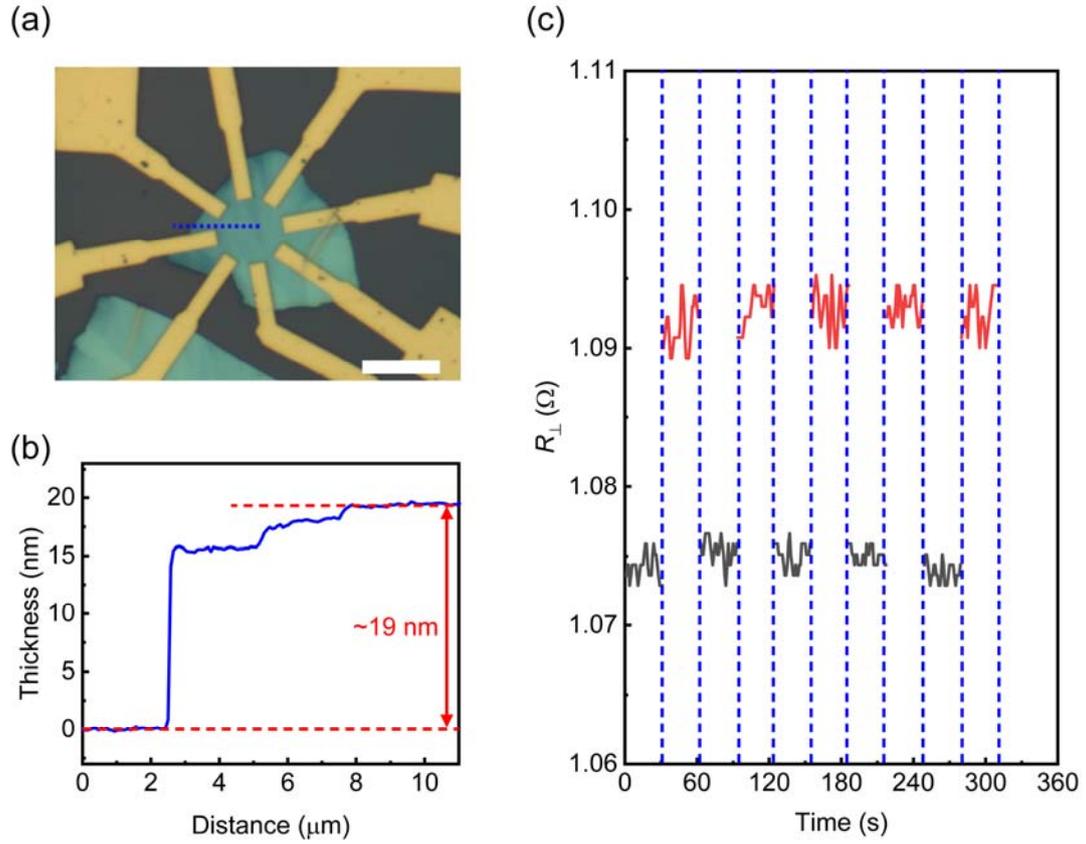

Fig. S1. Current-control of the helical antiferromagnetic order in another $Ni_{1/3}NbS_2$ device with ~19 nm thickness. (a) Optical image of the new thin device. The white scale bar is 10 µm. (b) The thickness is ~19 nm, as confirmed by the atomic force microscopy, tracing the blue dashed line in (a). (c) Current-control of the helical AFM order in this thin device.

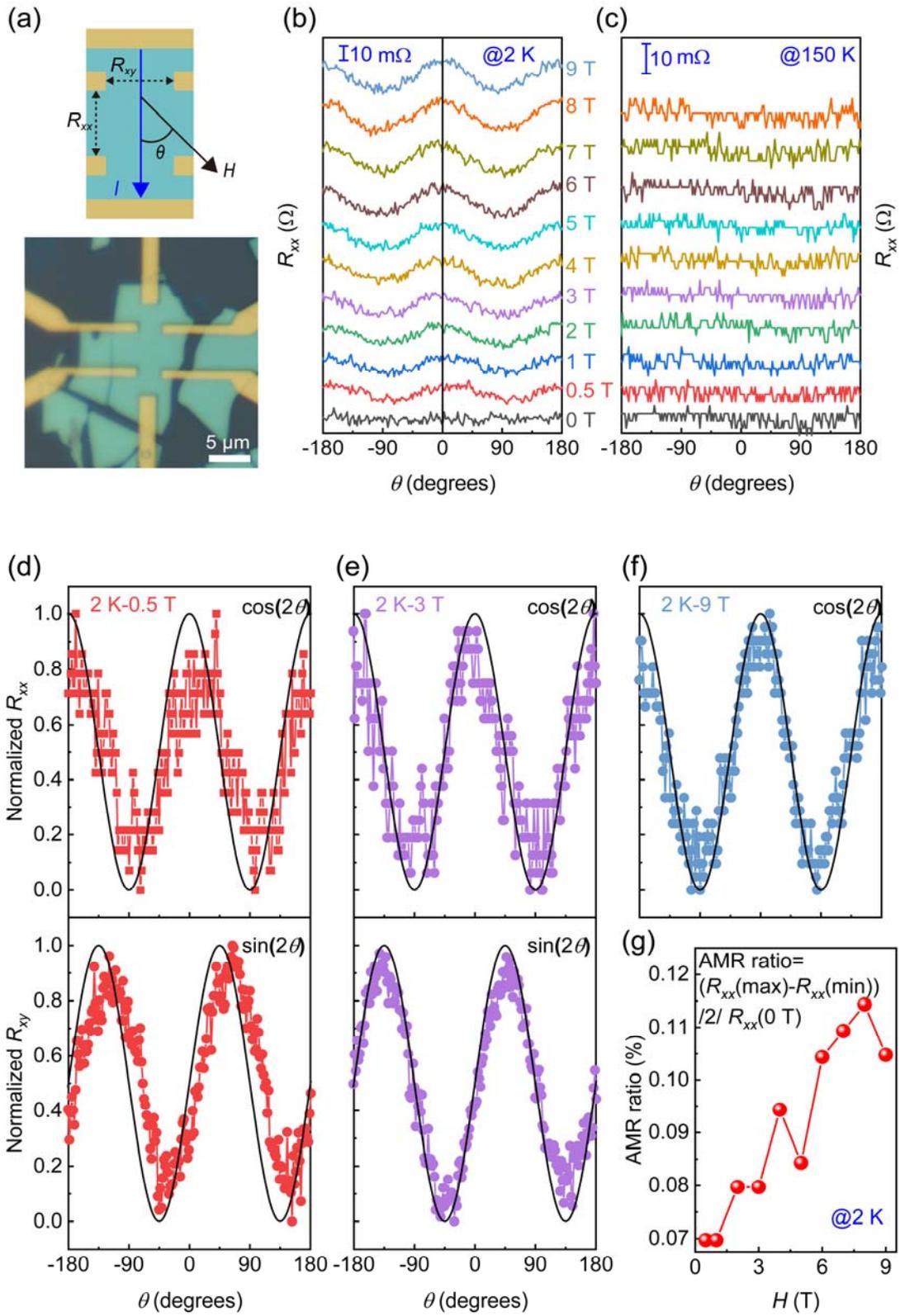

Fig. S2. AMR measurement. (a) AMR measurement schematic and optical image of the Ni$_{1/3}$NbS$_2$ device. The magnetic field is rotated in the sample plane with an angle $\theta$ to the current direction. (b) AMR result, i.e., the $R_{xx}$-$\theta$ curve under various magnetic fields at 2 K below the $T_{\text{Néel}}$ of 85 K, featuring clear oscillations. (c) $R_{xx}$-$\theta$ curve at 150 K above the $T_{\text{Néel}}$ of 85 K, showing a flat resistance background with no oscillations. (d-e) Normalized $R_{xx}$-$\theta$ and $R_{xy}$-$\theta$ curves at 2 K under 0.5 (d) and 3 T (e), respectively. $R_{xx}$, and $R_{xy}$ roughly follow the $\cos 2\theta$ and $\sin 2\theta$ behaviors (black curves). However, a close examination reveals that the experimental data deviates from the ideal sinusoidal curves, indicating the presence of an in-plane magnetic anisotropy. (f) Normalized $R_{xx}$-$\theta$ curve at 2 K under 9 T. It follows the theoretical line much more closely but still does not fully overlap with the ideal $\cos 2\theta$ curve (black curve). (g) AMR ratio, defined as $(R_{xx}(\text{max})-R_{xx}(\text{min}))/2/R_{xx}(0\text{ T})$. It rises with the increasing magnetic field, at least up to 9 T, but the AMR ratio remains tiny, around the order of 0.1%, which is why the measurement is difficult, and the obtained experimental data of the nanoflake suffers from noise.

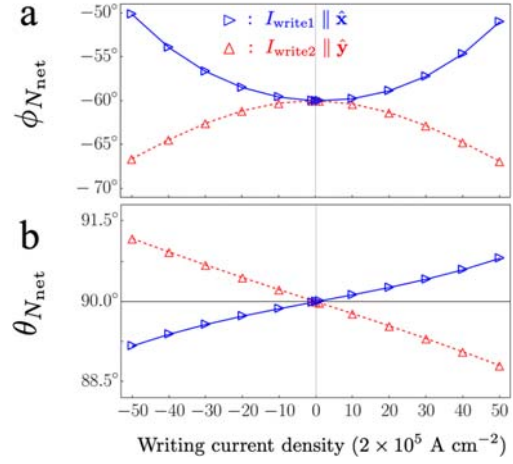

Fig. S3. Electrical control of the net Néel order. (a) Azimuthal and (b) polar angles for the net Néel order **N**$_\text{net}$ as a function of the writing current density. The equilibrium **N**$_\text{net}$ is set towards $\phi_{N_\text{net},0} = -60°$ to the $x$-axis within the $xy$-plane.

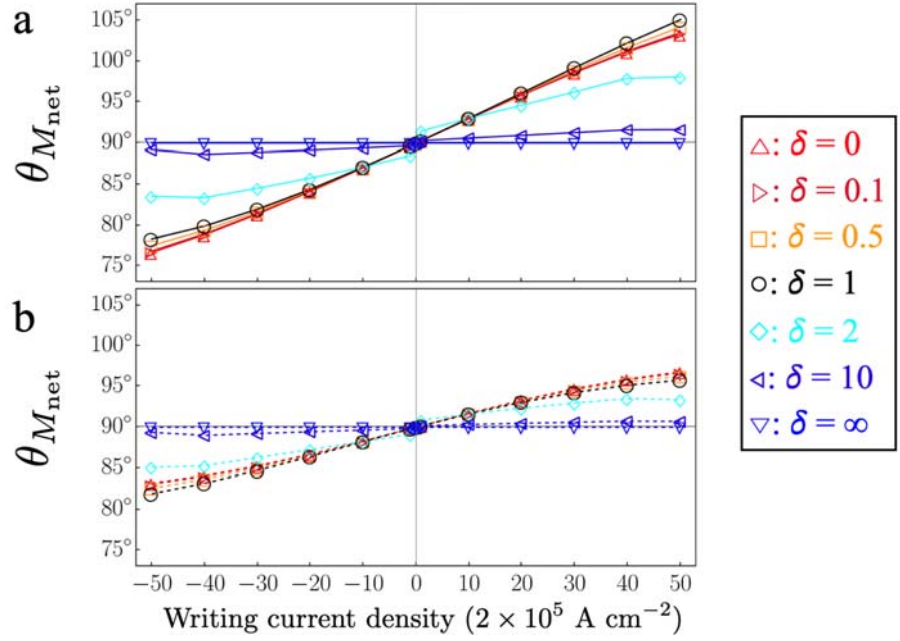

Fig. S4. Polar angles as a function of the writing current density by field-like and antidamping-like spin-orbit torques ratio $\delta$. In (a) and (b), the writing current points along the $x$ and $y$ axis, respectively. Note that the corresponding azimuthal angles have been presented in Figs. 2(f-g) in the main text.

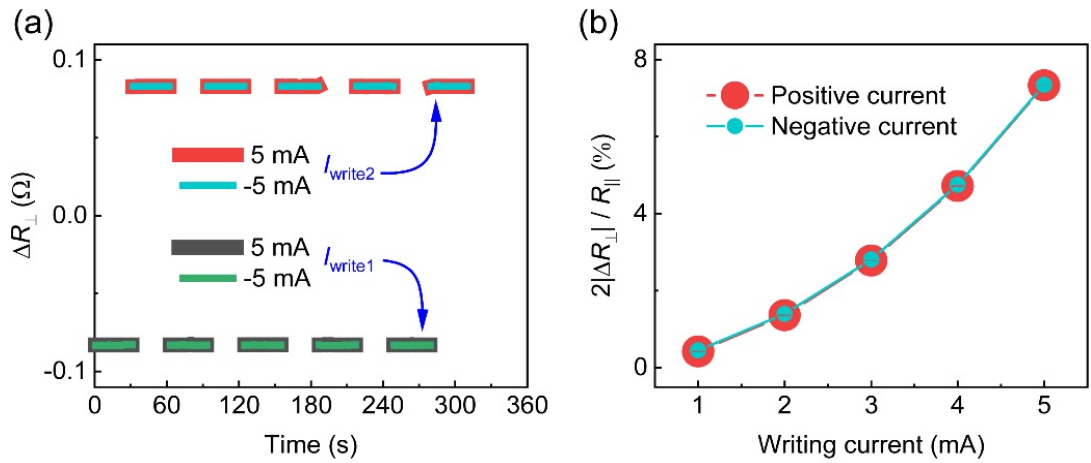

Fig. S5. Current polarization dependence. (a) Current-controlled resistance change of 5 mA and -5 mA for $I_{write1}$ and $I_{write2}$. The resistance change is almost the same. (b) Relative resistance changes as a function of the writing current, independent of the writing current polarization.

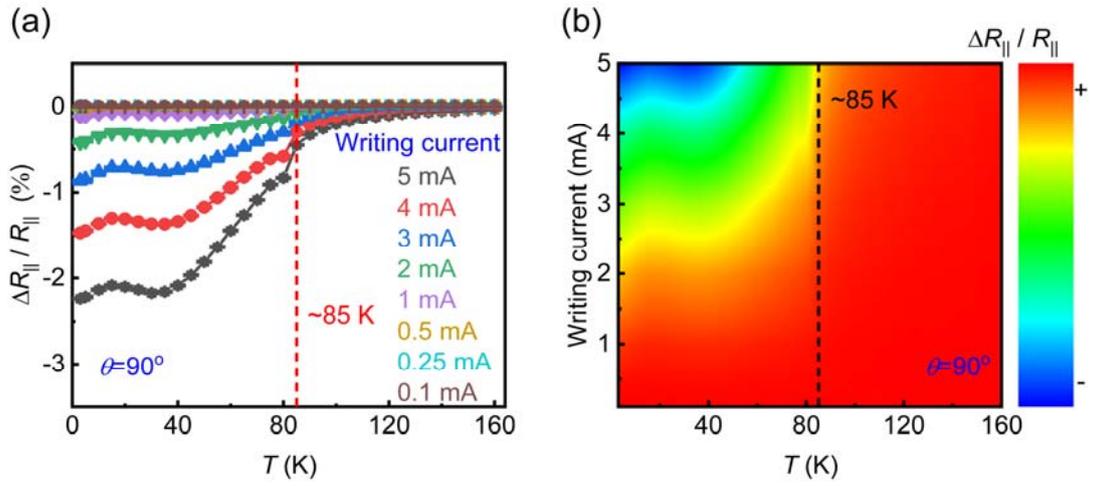

Fig. S6. Temperature dependence of the current-controlled resistance change for $\theta=90°$. (a) Relative resistance change $\Delta R_{\parallel}/R_{\parallel}$ as a function of temperature for $\theta=90°$. The red dashed line indicates the Néel temperature of ~85 K. (b) Contour-plot mapping of relative resistance change $\Delta R_{\parallel}/R_{\parallel}$ across the temperature-writing current parameter space for $\theta=90°$. The black dashed line indicates the Néel temperature of ~85 K.

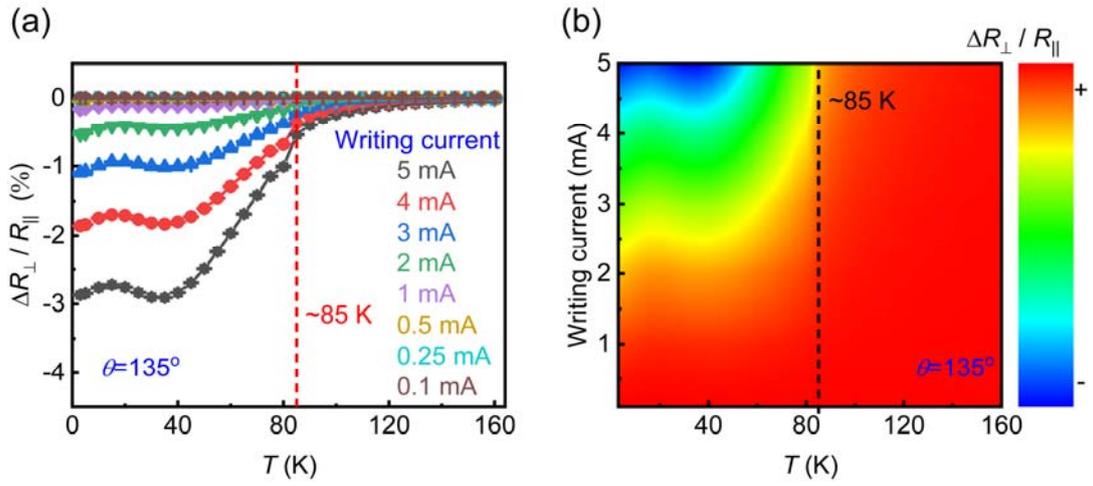

Fig. S7. Temperature dependence of the current-controlled resistance change for $\theta=135°$. (a) Relative resistance change $\Delta R_\perp/R_\parallel$ as a function of temperature for $\theta=135°$. The red dashed line indicates the Néel temperature of ~85 K. (b) Contour-plot mapping of relative resistance change $\Delta R_\perp/R_\parallel$ across the temperature-writing current parameter space for $\theta=135°$. The black dashed line indicates the Néel temperature of ~85 K.

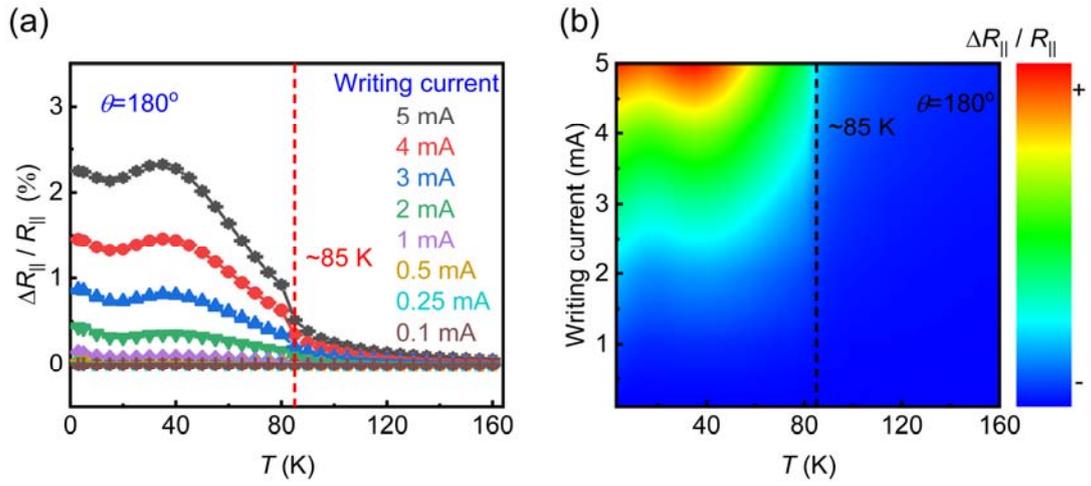

Fig. S8. Temperature dependence of the current-controlled resistance change for $\theta=180°$. (a) Relative resistance change $\Delta R_\parallel / R_\parallel$ as a function of temperature for $\theta=180°$. The red dashed line indicates the Néel temperature of ~85 K. (b) Contour-plot mapping of relative resistance change $\Delta R_\parallel / R_\parallel$ across the temperature-writing current parameter space for $\theta=180°$. The black dashed line indicates the Néel temperature of ~85 K.

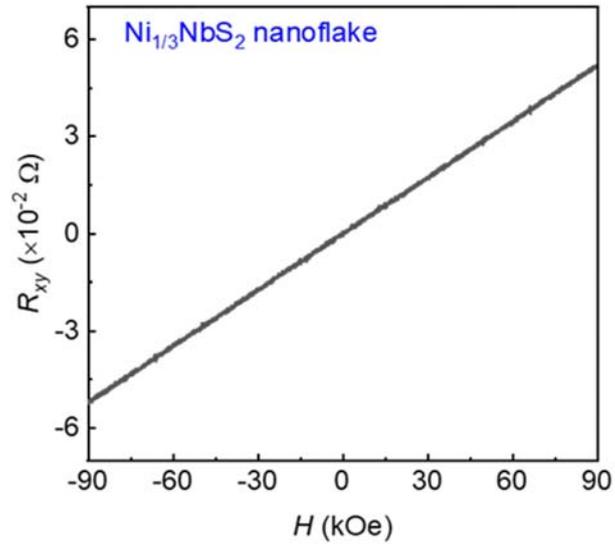

Fig. S9. $R_{xy}$-$H$ curve of the Ni$_{1/3}$NbS$_2$ nanoflake with magnetic field up to 9 T at 40 K below the Néel temperature. It shows a perfectly ordinary Hall effect with no hysteresis loop of anomalous Hall effect and no sudden magnetic transition up to 9 T, indicating the robustness of our helical antiferromagnet.

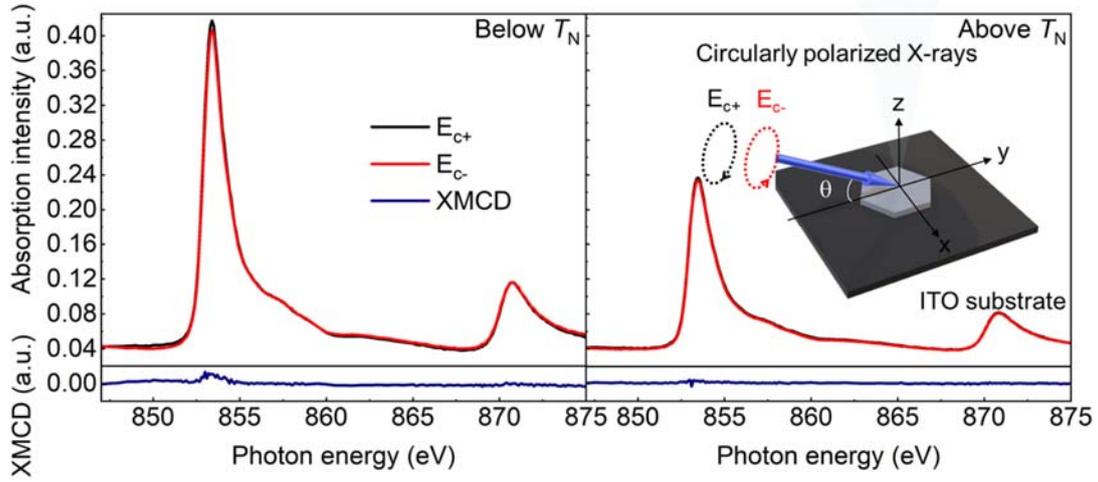

Fig. S10. X-ray magnetic circular dichroism (XMCD) response of a typical Ni$_{1/3}$NbS$_2$ nanoflake measured below and above $T_N$. The inset shows the measurement schematic with circularly polarized X-rays and an incident angle of 16°, performed under photoemission electron microscopy (PEEM) at the Paul Scherrer Institute (PSI). Below $T_N$ (55 K), the absorption peak for the right- and left-handed circularly polarized light exhibits a small but clear difference, represented by the XMCD plot derived from their subtraction. The XMCD, featuring a subtle peak/hump at the photon energy of 853 eV around the Ni L$_3$-edge, indicates the presence of weak **M**$_{net}$ in the Ni$_{1/3}$NbS$_2$ system. In contrast, no noticeable XMCD response is observed above $T_N$ (160 K), confirming the magnetic ordering origin for the XMCD below $T_N$. While the weak XMCD signal limits quantitative analysis, it provides evidence of the weak net ferromagnetic component in this layered helical antiferromagnet.

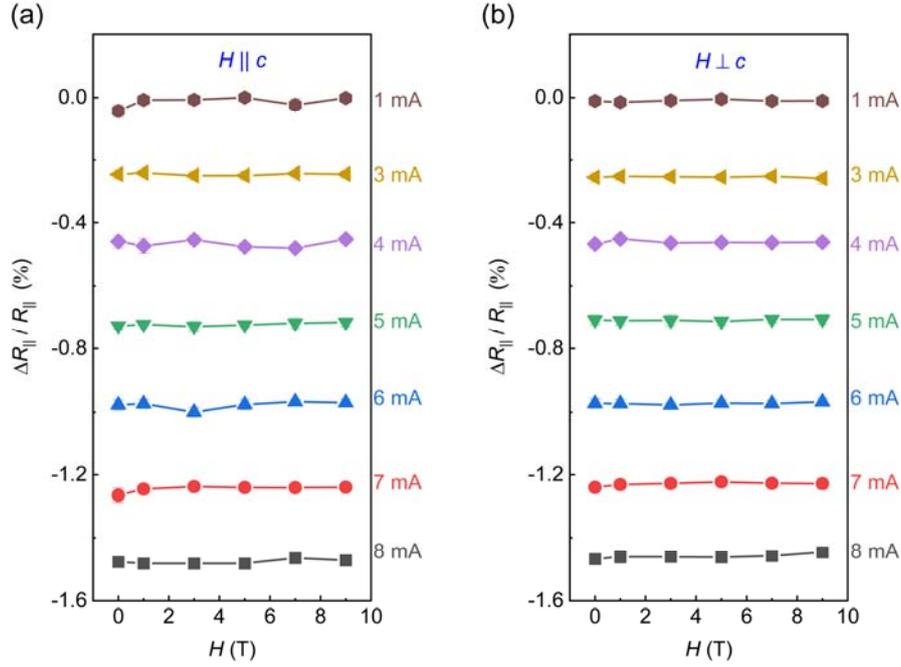

Fig. S11. Magnetic field dependence of helical antiferromagnet $Ni_{1/3}NbS_2$. (a-b), Relative resistance change $\Delta R_\parallel/R_\parallel$ of another nanodevice as a function of the magnetic field with various writing currents for $H \parallel c$ (a) and $H \perp c$ (b), respectively. $\Delta R_\parallel/R_\parallel$, *i.e.*, the antiferromagnetic switching shows weak magnetic field dependence in both cases, except for a slight decrease in magnitude at high magnetic fields. Such stability of resistance switching implies that the AFM memory can be read and written by current-driven SOT in the presence of strong magnetic field perturbations [12].

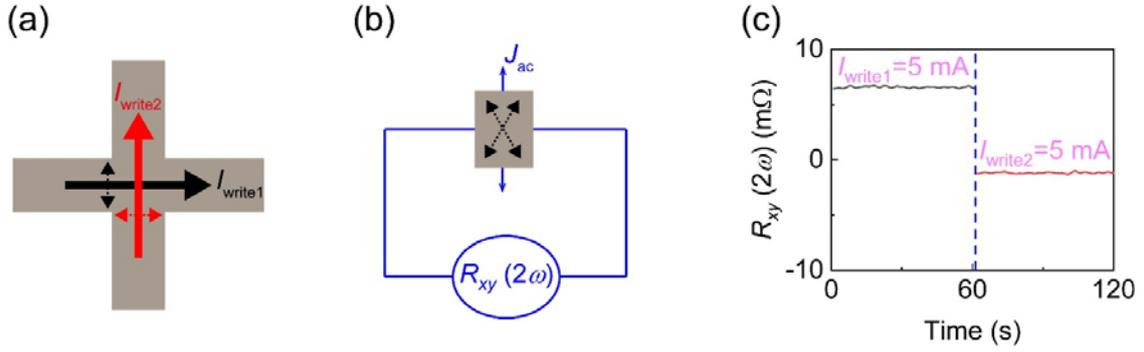

Fig. S12. Second harmonic measurement on helical antiferromagnet $Ni_{1/3}NbS_2$. (a) Expected spin reorientation trend by current-driven spin-orbit torque. Writing current tends to push the spin perpendicular to the charge current direction, as highlighted by the dashed double arrow. (b) Second harmonic measurement schematic: a.c. current is applied vertically in the sample plane, and the transverse Hall resistance of the second harmonic is monitored. For the $I_{write1}$ case, the spin-orbit torque by $I_{write1}$ pushes the spin vertically in (a); so when a.c. current is applied vertically in the same direction as the spins, it will cause the spin oscillation with corresponding resistance change and maximize the second harmonic signal. However, for the $I_{write2}$ case, the spin-orbit torque by $I_{write2}$ pushes the spin horizontally in (a); when a.c. current is applied vertically, it will not generate a second harmonic signal since the spins have already been in the favorable direction and will not be perturbed much by the a.c. current. (c) Second harmonic resistance $R_{xy}(2\omega)$ in our experiment as a function of time under the $I_{write1}$ and $I_{write2}$ cases. As expected from the SOT scenario and above analysis, $R_{xy}(2\omega)$ presents a significant value for the $I_{write1}$ case while it is nearly zero for the $I_{write2}$ case, which again validates the SOT scenario [13,14]. Note that our second harmonic results are similar to the electrically induced second harmonic response in the famous CuMnAs system [14].

**Supplemental References**

[1] H. Wu, H. Zhang, B. Wang, F. Gross, C. Y. Yang, G. Li, C. Guo, H. He, K. Wong, D. Wu, X. Han, C. H. Lai, J. Grafe, R. Cheng, and K. L. Wang, Current-induced Neel order switching facilitated by magnetic phase transition, Nat. Commun. **13**, 1629 (2022).

[2] S. Seki, X. Yu, S. Ishiwata, and Y. Tokura, Observation of skyrmions in a multiferroic material, Science **336**, 198 (2012).

[3] N. D. Khanh, T. Nakajima, X. Yu, S. Gao, K. Shibata, M. Hirschberger, Y. Yamasaki, H. Sagayama, H. Nakao, L. Peng, K. Nakajima, R. Takagi, T. H. Arima, Y. Tokura, and S. Seki, Nanometric square skyrmion lattice in a centrosymmetric tetragonal magnet, Nat. Nanotechnol. **15**, 444 (2020).

[4] H. Yoshimochi, R. Takagi, J. Ju, N. D. Khanh, H. Saito, H. Sagayama, H. Nakao, S. Itoh, Y. Tokura, T. Arima, S. Hayami, T. Nakajima, and S. Seki, Multistep topological transitions among meron and skyrmion crystals in a centrosymmetric magnet, Nat. Phys. **20**, 1001 (2024).

[5] S. S. Parkin, M. Hayashi, and L. Thomas, Magnetic domain-wall racetrack memory, Science **320**, 190 (2008).

[6] L. Thomas, R. Moriya, C. Rettner, and S. S. P. Parkin, Dynamics of Magnetic Domain Walls Under Their Own Inertia, Science **330**, 1810 (2010).

[7] S. Parkin and S. H. Yang, Memory on the racetrack, Nat. Nanotechnol. **10**, 195 (2015).

[8] Z. Luo, A. Hrabec, T. P. Dao, G. Sala, S. Finizio, J. Feng, S. Mayr, J. Raabe, P. Gambardella, and L. J. Heyderman, Current-driven magnetic domain-wall logic, Nature **579**, 214 (2020).

[9] M. Wu, T. Chen, T. Nomoto, Y. Tserkovnyak, H. Isshiki, Y. Nakatani, T. Higo, T. Tomita, K. Kondou, R. Arita, S. Nakatsuji, and Y. Otani, Current-driven fast magnetic octupole domain-wall motion in noncollinear antiferromagnets, Nat. Commun. **15**, 4305 (2024).

[10] N. L. Nair, E. Maniv, C. John, S. Doyle, J. Orenstein, and J. G. Analytis, Electrical switching in a magnetically intercalated transition metal dichalcogenide, Nat. Mater. **19**, 153 (2020).

[11] E. Maniv, N. L. Nair, S. C. Haley, S. Doyle, C. John, S. Cabrini, A. Maniv, S. K. Ramakrishna, Y. L. Tang, P. Ercius, R. Ramesh, Y. Tserkovnyak, A. P. Reyes, and J. G. Analytis, Antiferromagnetic switching driven by the collective dynamics of a coexisting spin glass, Sci. Adv. **7**, eabd8452 (2021).

[12] P. Wadley, B. Howells, J. Zelezny, C. Andrews, V. Hills, R. P. Campion, V. Novak, K. Olejnik, F. Maccherozzi, S. S. Dhesi, S. Y. Martin, T. Wagner, J. Wunderlich, F. Freimuth, Y. Mokrousov, J. Kunes, J. S. Chauhan, M. J. Grzybowski, A. W. Rushforth, K. W. Edmonds, B. L. Gallagher, and T. Jungwirth, Electrical switching of an antiferromagnet, Science **351**, 587 (2016).

[13] C. O. Avci, K. Garello, M. Gabureac, A. Ghosh, A. Fuhrer, S. F. Alvarado, and P. Gambardella, Interplay of spin-orbit torque and thermoelectric effects in ferromagnet/normal-metal bilayers, Phys. Rev. B **90**, 224427 (2014).

[14] J. Godinho, H. Reichlova, D. Kriegner, V. Novak, K. Olejnik, Z. Kaspar, Z. Soban, P. Wadley, R. P. Campion, R. M. Otxoa, P. E. Roy, J. Zelezny, T. Jungwirth, and J. Wunderlich, Electrically induced and detected Neel vector reversal in a collinear antiferromagnet, Nat. Commun. **9**, 4686 (2018).